\documentclass[preprint2,twocolumn,times,tighten]{aastex631}
\usepackage{graphicx}	
\usepackage{amsmath}	
\usepackage{amssymb}	
\let\tablenum\relax
\usepackage{siunitx}

\begin{document}

\title{Cosmic Ray Perpendicular Superdiffusion and Parallel Mirror Diffusion in a Partially Ionized and Turbulent Medium}

\email{yuehu@ias.edu; *NASA Hubble Fellow}

\author[0000-0002-8455-0805]{Yue Hu*}
\affiliation{Institute for Advanced Study, 1 Einstein Drive, Princeton, NJ 08540, USA }

\author{Siyao Xu}
\affiliation{Department of Physics, University of Florida, 2001 Museum Rd., Gainesville, FL 32611, USA}

\author{Alex Lazarian}
\affiliation{Department of Astronomy, University of Wisconsin-Madison, Madison, WI, 53706, USA}

\author{James M. Stone}
\affiliation{Institute for Advanced Study, 1 Einstein Drive, Princeton, NJ 08540, USA }
\affiliation{Department of Astrophysical Sciences, Princeton University, Princeton, NJ 08540, USA}

\author[0000-0003-3729-1684]{Philip F. Hopkins}
\affiliation{TAPIR, California Institute of Technology, Pasadena, CA 91125, US}


\begin{abstract}
Understanding cosmic ray (CR) diffusion in a partially ionized medium is both crucial and challenging. In this study, we investigate CR perpendicular superdiffusion and parallel transport in turbulent, partially ionized media using high-resolution 3D two-fluid simulations that treat ions and neutrals separately. We examine the influence of neutral-ion decoupling and the associated damping of turbulence on CR propagation in both transonic and supersonic conditions. Our simulations demonstrate that neutral-ion decoupling significantly damps velocity and magnetic field fluctuations at small scales, producing spectral slopes steeper than those of Kolmogorov and Burgers scaling. In supersonic turbulence, large-scale shock motion is not subject to damping and generates small-scale density enhancements. Moreover, the damping of magnetic field fluctuations substantially decreases pitch-angle scattering, which, however, only slightly affects the CR parallel mean free path $\lambda_\|$, due to the nonresonant mirror interactions of CRs. In the direction perpendicular to the mean magnetic field, we identify two regimes of the perpendicular superdiffusion of CRs: a diffusive regime ($\lambda_\|<L_{\rm inj}$, where $L_{\rm inj}$ is turbulence injection scale) with perpendicular separation of CR proportional to $t^{3/4}$, and a ballistic-like regime ($\lambda_\|>L_{\rm inj}$), with perpendicular separation scaling as $t^{3/2}$. At initially large pitch angles, the effects of magnetic mirroring—naturally arising in magnetohydrodynamic turbulence—become significant, enhancing the confinement of CRs and resulting in $\lambda_\|<L_{\rm inj}$, despite the damping effect. These results imply that large-pitch-angle CRs can be well confined in the cold ISM, such as molecular clouds.
\end{abstract}




\section{Introduction} \label{sec:intro}
Cosmic rays (CRs) comprise a substantial fraction of energy reservoir in galaxies, driving processes that range from large-scale outflows \citep{2008ApJ...674..258E,2016ApJ...816L..19G,2017ApJ...834..208R,2021MNRAS.501.3640H,2025arXiv250118696H} and multi-messenger emissions \citep{2013Sci...342E...1I,2018Sci...361..147I} to feedback within star-forming regions \citep{2008A&A...481...33J,2009Natur.462..770V,2023MNRAS.520.5126K}. While low-energy ($\sim$GeV) CRs are often emphasized for their capacity to ionize and heat dense gas \citep{2007ApJ...671.1736I,2020ApJ...903...77B}, the TeV–PeV population also exerts a critical influence on the partially ionized interstellar medium (ISM), which has a wide range of ionization fractions \citep{2011piim.book.....D}. These high-energy particles penetrate deeply into molecular clouds, impart momentum, and generate secondary particles (e.g., gamma-ray–producing pions) \citep{2013PhPl...20e5501Z,2013A&ARv..21...70B,2015ApJ...802..114K,2022MNRAS.516.3470H}. For instance, the detection of TeV gamma-ray and synchrotron radio emission from giant molecular clouds like Sgr~B2 and W43 indicates an enhanced local population of high-energy CRs \citep{2008A&A...481..401A,2024MNRAS.527.1275Y,2025SCPMA..6879502C}, suggesting that non-classical transport processes may be at work. This is because classical diffusion models, like the quasi-linear theory that treats the CR motions as a simple random walk \citep{1966ApJ...146..480J}, predict that CRs would not be confined to the dense clouds due to weak resonant scattering when damping of MHD turbulence is important. Therefore, understanding the transport properties of CRs in a partially ionized medium is essential for accurately characterizing nonthermal radiation.

A key determinant of CR transport is the nature of turbulent magnetic fields. Fluctuating fields introduce pitch-angle scattering and magnetic mirroring, which can affect the diffusion of CRs along field lines \citep{1966ApJ...146..480J,1969ApJ...155..777J,2002ApJ...578L.117Q,2021ApJ...923...53L}. A growing body of work has revealed that CR transport can deviate significantly from the classical diffusion paradigm in the presence of magnetohydrodynamic (MHD) turbulence \citep{2018PhRvL.121y5101C,2019ApJ...886..122C,2022MNRAS.517.5413H,2022MNRAS.514..657K,2022FrASS...9.0900B,2023FrASS..1054760L,2023JPlPh..89e1701L,2023MNRAS.525.4985K,2024ApJ...975...65Z,2025FrASS..1111076M}. In particular, turbulent magnetic field lines' superdiffusion \citep{LV99} induces superdiffusive transport perpendicular to mean magnetic field \citep[e.g.,][]{2008ApJ...673..942Y,2013ApJ...779..140X,2014ApJ...784...38L,2022MNRAS.512.2111H}. \cite{2021ApJ...923...53L} combine the effects of CR bouncing in magnetic mirrors \citep{1973ApJ...185..153C} results in parallel propagation that strongly depends on the CR pitch angle, i.e., in the process termed "mirror diffusion" \citep{2021ApJ...923...53L,2023ApJ...957...97X,2025ApJ...988..269B,2025A&A...699A.317X}. However, most earlier studies assume a fully ionized plasma or adopt single-fluid approximations. In molecular cloud, on the contrary, the gas is cold ($\sim$10–100 K) and only weakly ionized, resulting in additional physical processes such as ion-neutral collisions and ion-neutral collisional damping of turbulence \citep[e.g.,][]{1986MNRAS.220..133D,1996ApJ...465..775B,2011MNRAS.415.3681T,2015ApJ...810...44X}. These effects can alter both the local turbulence spectrum and CR scattering rates \citep{2010MNRAS.406.1201T,2010ApJ...720.1612M,2015ApJ...805..118B,2024MNRAS.527.3945H}. Consequently, a two-fluid (ion + neutral) description is critical to fully capture the physics that governs CR propagation in these environments. By incorporating two-fluid effects, one can better connect observations—such as locally enhanced gamma-ray emission in molecular clouds—to the underlying plasma and magnetic field conditions.

Despite the high numerical cost to resolve extremely large Alfv\'en speeds in the neutral-ion decoupled status, the two-fluid approach is necessary, as earlier work has shown that neutral-ion decoupling can significantly modify the turbulent energy spectra of ions and neutrals \citep{2010MNRAS.406.1201T,2010ApJ...720.1612M,2015ApJ...805..118B}. In particular, \cite{2024MNRAS.527.3945H} found that large density fluctuations in ions and neutrals can drive substantial spatial variations in the ionization fraction and local neutral-ion coupling status. Before reaching the locally smallest decoupling scale, i.e., the fully decoupled regime, neutrals can remain partially coupled with ions in an intermediate coupled regime. In the intermediately coupled and fully decoupled regimes, turbulence can be damped, resulting in steeper kinetic and magnetic field spectra than the familiar Kolmogorov profile, which profoundly influences CR propagation.

In this work, we extend existing theories of magnetic turbulence and CR transport by incorporating essential two-fluid physics of neutral-ion decoupling and the associated collisional damping of turbulence. We perform two-fluid MHD simulations using the AthenaK code \citep{2024arXiv240916053S}, varying the ionization fraction and coupling strength to explore different regimes of ion-neutral interaction. By examining the resulting CR transport properties, we aim to clarify how partial ionization modifies turbulence and, consequently, CR transport in dense interstellar environments.

This paper is organized as follows. \S~\ref{sec:theory} provides a brief review of the theoretical framework for CR perpendicular superdiffusion and parallel mirror diffusion in a magnetized, partially ionized medium. \S~\ref{sec:data} describes the 3D two-fluid simulations and numerical approach of modeling CR propagation used in this study. In \S~\ref{sec:results}, we analyze the statistics of turbulence and CR diffusion in the partially ionized medium at different degrees of neutral-ion coupling. We conclude with a summary of our main findings in \S~\ref{sec:conclusion}.

\section{Theoretical consideration}
\label{sec:theory}
\subsection{Decoupling of neutrals from ions}
The coupling between ions and neutrals is commonly characterized by two collisional frequencies. The neutral–ion collision frequency is defined as $\nu_{ni} = \gamma_{\rm d} \rho_i=\gamma_{\rm d}\xi_i(\rho_i+\rho_n)$ while the ion–neutral collision frequency is given by $\nu_{in} = \gamma_{\rm d} \rho_n$, respectively \citep{1992pavi.book.....S}. Here $\gamma_{\rm d}$ represents the drag coefficient that couples ions and neutrals through the collisional exchange of momentum, while $\xi_i=\rho_i/(\rho_i+\rho_n)$ is the ionization fraction.  $\rho_i$ and $\rho_n$ are ion mass density and neutral mass density, respectively. Decoupling of neutrals from ions begins when the rate at which turbulent energy cascades becomes comparable to $\nu_{ni}$. Because $\nu_{in}$ greatly exceeds $\nu_{ni}$ in weakly ionized media, ions separate from neutrals on much smaller scales than those marking the decoupling of neutrals; hence, our focus here is primarily on the neutral–ion decoupling process.

Early linear studies proposed that neutrals decouple from Alfv\'enic oscillations at a characteristic wavenumber $k_{{\rm dec},\parallel}$, determined by setting the Alfv\'en wave frequency equal to the neutral–ion collision frequency \citep{1969ApJ...156..445K,1992pavi.book.....S}: 
\begin{equation}
\label{eq.cp}
k_{{\rm dec},\parallel} v_A=\nu_{ni}.
\end{equation}
Here, the subscript "$\parallel$" designates the component along the magnetic field, and $v_A$ is the Alfv\'en speed. 

Although MHD turbulence was approximated as a superposition of linear waves in some earlier studies \citep{1999ApJ...520..204G}, it is now recognized as an inherently non-linear process. In particular, within the strong turbulence regime under a critical balance condition, any Alfv\'en wave behavior is short-lived, persisting for a wave period \citep{GS95,LV99}. With the dynamically coupled turbulent mixing motion in the perpendicular direction and the wave-like motion in the parallel direction, the cascade of the MHD turbulence is mainly along the perpendicular direction, and the scale-dependent anisotropy of MHD turbulence develops (see \citealt{xu2019study} for a review). For Alfv\'enic turbulence \footnote{For a more in-depth discussion on neutral-ion decoupling scales of the three MHD modes (Alfv\'en, fast, and slow), see \cite{2015ApJ...810...44X,2016ApJ...826..166X}.}, which typically dominates the turbulent energy budget \citep{2002ApJ...564..291C,HLX21a}, this anisotropy implies that the decoupling length is itself anisotropic—the parallel component can exceed the perpendicular one, so that the latter is smaller than earlier estimates \citep{2015ApJ...810...44X,2016ApJ...826..166X,2024MNRAS.527.3945H}. The perpendicular decoupling scale in strong MHD turbulence regimes is expressed as \citep{2015ApJ...810...44X}:
\begin{equation}
\label{eq:cp_perp}
\begin{aligned}
        k_{{\rm dec},\bot}&=
        \begin{cases}
        \nu_{\rm ni}^{3/2}L_{\rm inj}^{1/2}v_{\rm inj}^{-3/2},~~~M_A>1\\
        \nu_{\rm ni}^{3/2}L_{\rm inj}^{1/2}v_{\rm inj}^{-3/2}M_A^{-1/2},~~~M_A<1\\
        \end{cases},
\end{aligned}
\end{equation}
where $L_{\rm inj}$ and $v_{\rm inj}$ are the injection scale and injection velocity of turbulence, respectively. $M_A = v_{\rm inj}/v_A$ is the Alfv\'en Mach number.

\begin{figure*}[p]
  \centering
  \gridline{
    \fig{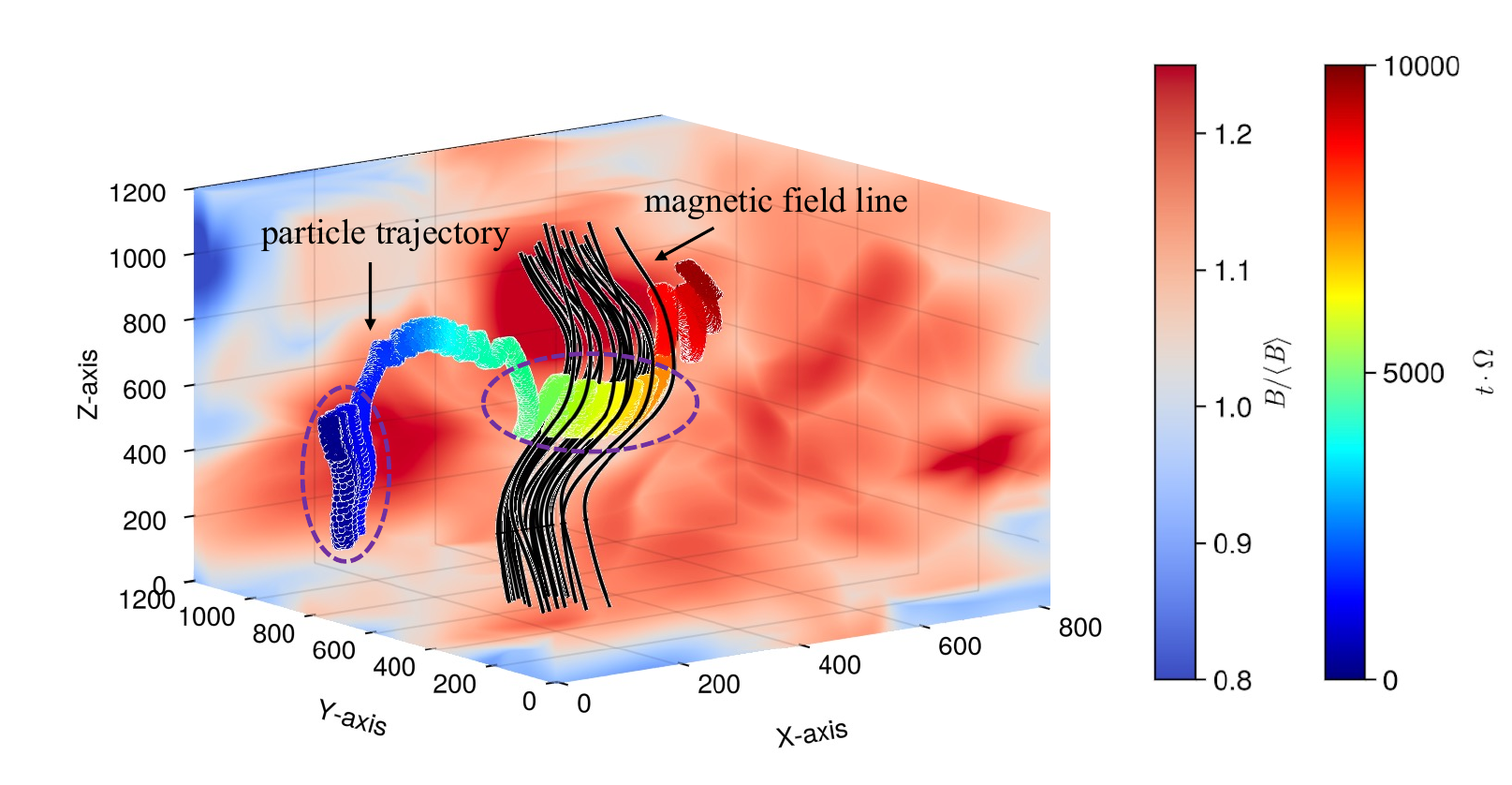}{0.99\textwidth}{(a) particle trajectory colored with time}
  }
  \vspace{-1.5em}
  \gridline{
    \fig{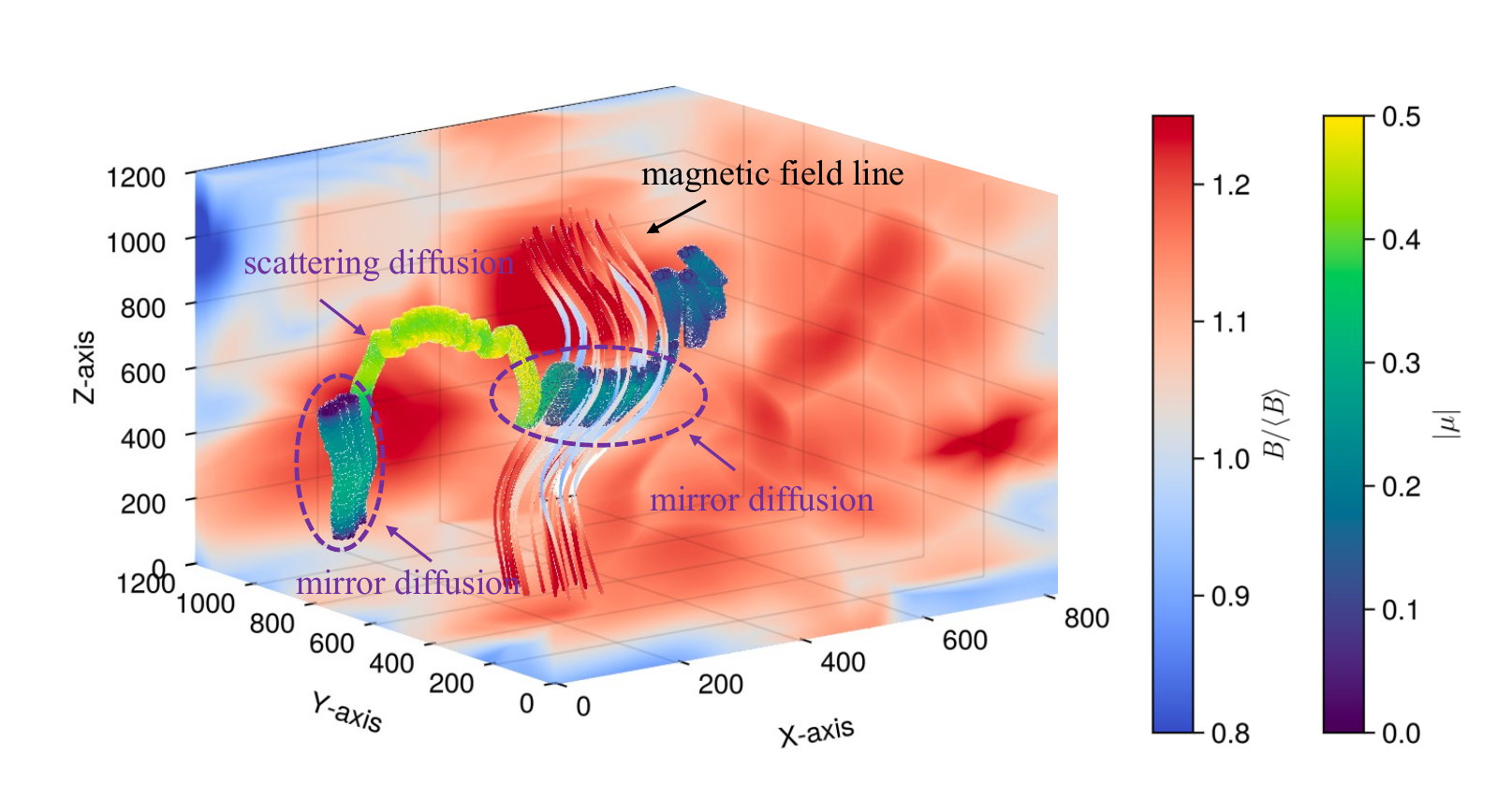}{0.99\textwidth}{(b) particle trajectory colored with $|\mu|$}
  }
  \caption{An example of the scattering diffusion and mirror diffusion. A particle with a Larmor radius of 10 cells and an initial pitch angle cosine of 0.25 is ejected into the simulation A4 ($M_s=1.24$ and $M_A=1.21$). The particle trajectory is plotted in the 3D volume rendering cube of normalized magnetic field strength. Panel (a): The trajectory is colored to show the time,  increasing from blue to red. Panel (b): The trajectory is colored to show $|\mu|$, where $\mu$ is the cosine pitch angle, increasing from dark to yellow. Magnetic field lines are colored to show the normalized magnetic field strength. In the regimes dominated by a magnetic field mirror, the particle is reflected back and forth by multiple mirrors, while undergoing superdiffusion perpendicular to the mean magnetic field. It can reenter the scattering diffusion regime due to the gradual stochastic change of pitch angle.}
  \label{fig:illu}
\end{figure*}

\subsection{Neutral-ion collisional damping of MHD turbulence}
\cite{2024MNRAS.527.3945H} further revealed that, in compressible MHD turbulence, large density fluctuations in both ions and neutrals lead to spatial variations in the Alfv\'en speed and ionization fraction. These variations cause the parallel and perpendicular neutral–ion decoupling scales to span over an extended range. Because of the spatial variation of the decoupling scales, the overall coupling status of neutrals with ions can be categorized into three regimes.

\textbf{Strongly coupled regime:} In the strongly coupled regime, where scales are larger than the maximum parallel decoupling length, i.e., $k<k_{\rm dec,\|}$, ions and neutrals move in unison, effectively forming a single fluid that carries the MHD turbulence. Both species share and sustain the cascading turbulent energy without any noticeable separation in their dynamics.

\textbf{Intermediate coupling:} 
The intermediately coupled regime corresponds to intermediate scales where the neutral and ions can be either locally coupled or locally decoupled, depending on the local physical conditions. Although ions continue to experience frequent collisions with nearby neutrals—extending down to the ion–neutral decoupling scale—the local decoupling of neutrals leads to a significant modification of the ion dynamics. The MHD turbulence in ions is strongly damped by the friction forces arising from these collisions. In this regime, the hydrodynamic turbulence in neutrals is also modified locally \citep{2024MNRAS.527.3945H}.

\textbf{Fully decoupled regime:} In the fully decoupled regime, where the scales are below the minimum perpendicular decoupling scale, i.e., $k>k_{\rm dec,\bot}$, the behavior of the two components diverges markedly. Ions remain subject to frequent collisions with the surrounding neutrals, which results in the damping of their MHD turbulent cascade. Conversely, as neutrals fully decouple from ions, neutrals develop an independent hydrodynamic cascade. This decoupling of neutrals from ions leads to a clear separation in the dynamical evolution: while the ion turbulence remains damped by collisional friction, the neutrals evolve under a purely hydrodynamic regime.

The damping of MHD turbulence leads to a marked reduction in fluctuations of density, velocity, and especially the magnetic field. These diminished fluctuations in the magnetic field curtail the efficiency of pitch angle scattering for CRs and also affect the properties of magnetic mirrors, thereby altering CR's transport.

\begin{table*}[h!]
	\centering
 \begin{tabular}{ | c | c | c | c | c | c | c | c | c | c | c |}
		\hline
		Run & $M_s$ & $M_A$ &  $\frac{\gamma_{\rm d}\rho_i}{v_{\rm inj}/L_{\rm inj}}$ & $k_{\rm dec, \parallel}\frac{L_{\rm box}}{2\pi}$ & $k_{\rm dec, \bot}\frac{L_{\rm box}}{2\pi}$ &  $l_{\rm dec, \parallel}$ [pc] & $l_{\rm dec, \bot}$ [pc] & $l_{\rm dis}$ [pc] & $L_{\rm box}$ [pc] & Coupling status \\ \hline \hline
		A1 & 1.00 & 0.87 &  $5\times10^3$ & $10^4$ & $7\times10^5$ & $10^{-6}$ & $1.4\times10^{-8}$ & $10^{-4}$ & $10^{-2}$ & strong coupling \\
		A2 & 0.88 & 0.78 & $5\times10^2$ & $10^3$ & $2\times10^4$ & $10^{-5}$ & $5\times10^{-7}$ & $10^{-4}$ & $10^{-2}$ & coupling  \\
            A3 & 0.82 & 0.74 & $5\times10^1$ & $10^2$ & $7\times10^2$ & $10^{-4}$ & $1.4\times10^{-5}$ & $10^{-4}$ & $10^{-2}$ & intermediate coupling \\
            A4 & 1.24 & 1.21 &  $5\times10^0$ & $10^1$ & $2\times10^1$ & $10^{-3}$ & $5\times10^{-4}$ & $10^{-4}$ & $10^{-2}$ & decoupling \\ \hline
		B1 & 9.49 & 0.88 &  $5\times10^3$ & $10^4$ & $7\times10^5$ & $10^{-3}$ & $1.4\times10^{-5}$ & $10^{-1}$ & $10^1$ & strong coupling  \\
		B2 & 9.55 & 0.89 &  $5\times10^2$ & $10^3$ & $2\times10^4$ & $10^{-2}$ & $5\times10^{-4}$ & $10^{-1}$ & $10^1$ & coupling \\
            B3 & 9.60 & 0.91 &  $5\times10^1$ & $10^2$ & $7\times10^2$ & $10^{-1}$ & $1.4\times10^{-2}$ & $10^{-1}$& $10^1$ & intermediate coupling \\
            B4 & 11.05 & 1.09 &  $5\times10^0$ & $10^1$ & $2\times10^1$ & $10^{0}$ & $5\times10^{-1}$ & $10^{-1}$ & $10^1$ & decoupling  \\
        \hline
	\end{tabular}
	\caption{\label{tab:sim} The setup of two-fluid simulations at a resolution of $1200^3$ is summarized here. The sonic Mach number ($M_s$) and Alfv\'enic Mach number ($M_A$) represent the instantaneous RMS values at each snapshot. The plasma compressibility is given by $\beta = 2(M_A/M_s)^2$. The ratio of the neutral-ion collisional frequency to the eddy turnover rate at injection scale is expressed as $(\gamma_{\rm d} \rho_i) / (v_{\rm inj} / L_{\rm inj})$, which characterizes the coupling strength at injection scale. The theoretically expected parallel and perpendicular components of the neutral-ion decoupling wavenumber are denoted as $k_{\rm dec, \parallel}$ and $k_{\rm dec, \bot}$, respectively, while the corresponding neutral-ion decoupling scales are given by $l_{\rm dec, \parallel}$ and $l_{\rm dec, \bot}$. The numerical dissipation scale is represented by $l_{\rm dis}$. 
    }
\end{table*}

\subsection{CRs' perpendicular superdiffusion induced by perpendicular superdiffusion of magnetic field lines}

The perpendicular superdiffusion of CRs depends on their parallel diffusion behavior. It can be separated into two regimes based on the relationship between their mean free path $\lambda_\parallel$ (parallel to the magnetic field) and the injection scale of turbulence $L_{\rm inj}$.

\textbf{(1) When $\lambda_\parallel>L_{\rm inj}$:} the scattering and mirroring of CRs are negligible, and they move along magnetic field lines in a ballistic manner. The divergence of magnetic field lines primarily determines their diffusion perpendicular to the mean magnetic field. Specifically, the magnetic field lines separate superdiffusively in space \citep{LV99}:
\begin{equation}
\begin{aligned}
    \label{eq.B_str}
    \langle z^2\rangle\sim \frac{x^3}{27L_{\rm inj}}M_A^{4},~~{ M_A\le 1}, \\
    \langle z^2\rangle\sim \frac{x^3}{27L_{\rm inj}}M_A^{3},~~{ M_A> 1},
\end{aligned}
\end{equation}
where $x$ is the distance traced along magnetic field lines that are initially close to each other\footnote{A peculiar feature of this law is that for sufficiently large $x$, the line separation can be independent of the initial line separation as discussed in \cite{2011ApJ...743...51E}.} but larger than the damping scale, while $\sqrt{\langle z^2\rangle}$ is the averaged separation of magnetic field lines in the perpendicular direction gained as the result of this tracing. Note that the dependence on $M_A$ is different for sub-Alfv\'enic and super-Alfv\'enic turbulence. 

The magnetic field superdiffusion imprints its law on the CRs moving along magnetic field lines \citep{2013ApJ...779..140X}. Those exhibit superdiffusion in the direction perpendicular to the mean magnetic field \citep{2008ApJ...673..942Y, 2014ApJ...784...38L}. This superdiffusion in space is broadly related to the Richardson dispersion of magnetic field lines in time \cite{2013Natur.497..466E}, but is not equivalent to it. 
The relationship between the two can be exemplified by considering CR moving ballistically along the magnetic field lines with a negligible change in pitch angle and thus a constant parallel velocity. After time $t$, the CR traveled a distance of $x\sim c\mu t$ along the magnetic field line, where $c$ is the speed of light and $\mu$ is the pitch angle cosine. The CR's perpendicular separation can be expressed as:
\begin{equation}
\begin{aligned}
    \label{eq.cr_str}
    \langle z^2\rangle\sim \frac{(c\mu t)^3}{27L_{\rm inj}}M_A^{4},~~{ M_A\le 1}, \\
    \langle z^2\rangle\sim \frac{(c\mu t)^3}{27L_{\rm inj}}M_A^{3},~~{ M_A> 1},
\end{aligned}
\end{equation}
in which the functional dependence of $\sqrt{\langle z^2\rangle}$ on $t$ formally coincides with the law of the Richardson dispersion of magnetic field lines. However, the process described above is not the advection of magnetic field lines by turbulence described by Richardson diffusion, but the stochasticity of turbulent field lines due to their turbulent reconnection \citep{LV99}. Indeed, in our example above, the magnetic field lines are assumed to be static.

Note that Eq.~\ref{eq.cr_str} is valid for strong turbulence, specifically for scales smaller than $l_{\rm tr}=L_{\rm inj}M_A^2$ when $M_A\le1$ or $l_A=L_{\rm inj}M_A^{-3}$ when $M_A>1$. In the weak turbulence regime (i.e., scales larger than $l_{\rm tr}$ or $l_A$), turbulence is hydrodynamic-like rather than anisotropic-eddy-like. In this regime, the separation of magnetic field lines follows a random walk, and the CRs' perpendicular separation can be expressed as:
\begin{equation}
\begin{aligned}
    \label{eq.cr_weak}
    \langle z^2\rangle \sim \frac{c\mu tL_{\rm inj}}{3} M_A^4,~~{ M_A\le 1},\\
    \langle z^2\rangle \sim \frac{c\mu tL_{\rm inj}}{3} M_A^{-3},~~{ M_A> 1},\\
\end{aligned}
\end{equation}
which suggests a normal perpendicular diffusion of CRs at large scales. 
For a detailed derivation of these equations, see \cite{2008ApJ...673..942Y,2014ApJ...784...38L,2022MNRAS.512.2111H}. The transition from small-scale ($<L_{\rm inj}$) perpendicular superdiffusion to large-scale ($>L_{\rm inj}$) perpendicular normal diffusion has been numerically tested by \cite{2023ApJ...959L...8Z}.

\textbf{(2) When $\lambda_\parallel<L_{\rm inj}$:} the pitch-angle scattering and mirroring of CRs along magnetic field lines are non-negligible, and their propagation becomes diffusive rather than ballistic, as numerically tested in \citep{2022MNRAS.512.2111H,2023ApJ...959L...8Z}. 

Magnetic mirrors created by changes of magnetic field strength were assumed to trap CR with sufficiently large pitch angles, preventing them from further propagation unless their pitch angle decreases through resonant scattering \cite{1973ApJ...185..153C}. However, a recent study in \cite{2021ApJ...923...53L} suggested that in MHD turbulence, the perpendicular superdiffusion of turbulent magnetic field lines prevents CRs from being trapped between magnetic mirrors, as the return of CR to the initial mirror along the same field line is improbable, due to the field line stochasticity. Instead, CRs bounce among different magnetic mirrors while undergoing superdiffusion perpendicular to the mean field, leading to a diffusive motion along the magnetic field. This new parallel diffusion mechanism is termed "mirror diffusion" \citep{2021ApJ...923...53L}. According to \cite{2021ApJ...923...53L}, the transition between the scattering and mirror diffusion regimes occurs at a critical cosine of pitch angle $\mu_c$, corresponding to the balance between the scattering and mirroring rates. For CRs with a pitch angle cosine $\mu$ greater than $\mu_c$, parallel diffusion is governed primarily by gyroresonant pitch angle scattering, resulting in scattering diffusion; conversely, for $\mu$ less than $\mu_c$, mirror diffusion prevails. The mirror interactions of CRs are strong, as they reverse the CRs' motion. Thus, mirror diffusion is typically slower than that induced by resonance scattering \cite{2021ApJ...923...53L}. An example of the scattering diffusion and the mirror diffusion is given in Fig.~\ref{fig:illu}.

A particle with a Larmor radius of 10 cells and an initial pitch angle cosine of 0.25 is injected into simulation A4 (with $M_s=1.24$ and $M_A=1.21$). Initially, the particle enters the mirroring-dominated diffusion regime due to its large pitch angle, which causes it to bounce back and forth as it interacts with multiple mirrors. Over time, the particle gradually decreases its pitch angle due to scattering. Then, the particle parallel transport becomes dominated by scattering diffusion and later again undergoes the mirror diffusion due to the stochastic change of pitch angle. In the mirror diffusion regions, the cosine of pitch angle is quasiperiodic, while in the scattering diffusion regions, the cosine of pitch angle is close to constant. Similar effects have been seen in \cite{2023JPlPh..89e1701L}. Both scattering and mirror diffusion can contribute to the CR parallel diffusion behavior, and thus further affect the perpendicular superdiffusion behavior. If $\lambda_\parallel<L_{\rm inj}$, the CR traveled along the magnetic field line by a distance of $x^2\sim D_\parallel t$, the thus perpendicular superdiffusion can be expressed as follows \citep{2014ApJ...784...38L,2022MNRAS.512.2111H,2023FrASS..1054760L}:
\begin{equation}
\begin{aligned}
\label{eq.cr2}
    \langle z^2\rangle\sim \frac{(D_\|t)^{3/2}}{27L_{\rm inj}}M_A^{4},~~{ M_A\le 1}\\
    \langle z^2\rangle\sim \frac{(D_\|t)^{3/2}}{27L_{\rm inj}}M_A^{3},~~{ M_A> 1}
\end{aligned}
\end{equation}
where $D_\parallel$ is the parallel diffusion coefficient. The transport of CRs is thus highly sensitive to the value of their mean free path $\lambda_\parallel$, which is influenced by magnetic field fluctuations. 

\textbf{Strongly coupled regime:} In the strongly coupled regime, ions and neutrals together carry MHD turbulence through sufficient collisions and energy transfer. When CRs' Larmor radius is larger than the maximum parallel decoupling scale, they experience more significant magnetic field fluctuations than in the other decoupling regimes. The large fluctuations and thus higher scattering efficiency can cause a fast transition between scattering diffusion and mirror diffusion. Their perpendicular superdiffusion follows the scaling described by Eq.~\ref{eq.cr_weak}, with $\sqrt{\langle z^2\rangle}\propto t^{3/4}$.

\textbf{Intermediate coupling and fully decoupled regimes:} 
In the intermediate or fully decoupled regimes, the damping of MHD turbulence occurs, and magnetic field fluctuations are reduced, leading to an increase in $\lambda_\parallel$. In this regime, CRs experience insignificant scattering, and the superdiffusive behavior of magnetic field lines may be altered, at scales smaller than the maximum parallel decoupling scales. If the initial perpendicular separation of CRs is smaller than the maximum parallel decoupling scale, their separation will grow more slowly before entering the perpendicular superdiffusion regime described by Eq.~\ref{eq.cr_str}, with $\sqrt{\langle z^2\rangle}\propto t^{3/2}$. This superdiffusion is driven by the Richardson diffusion of magnetic field lines. The transition between these regimes underscores the critical role of ion–neutral coupling in regulating CR transport behavior. 

In addition, the damping of MHD turbulence also erases small-scale magnetic field mirrors. CRs can be captured by large-scale magnetic field mirrors. The large spatial separation within the mirrors could be sufficient for CRs entering the perpendicular superdiffusion regime and undergoing mirror diffusion.

\begin{figure*}[p]
\centering
\includegraphics[width=0.99\linewidth]{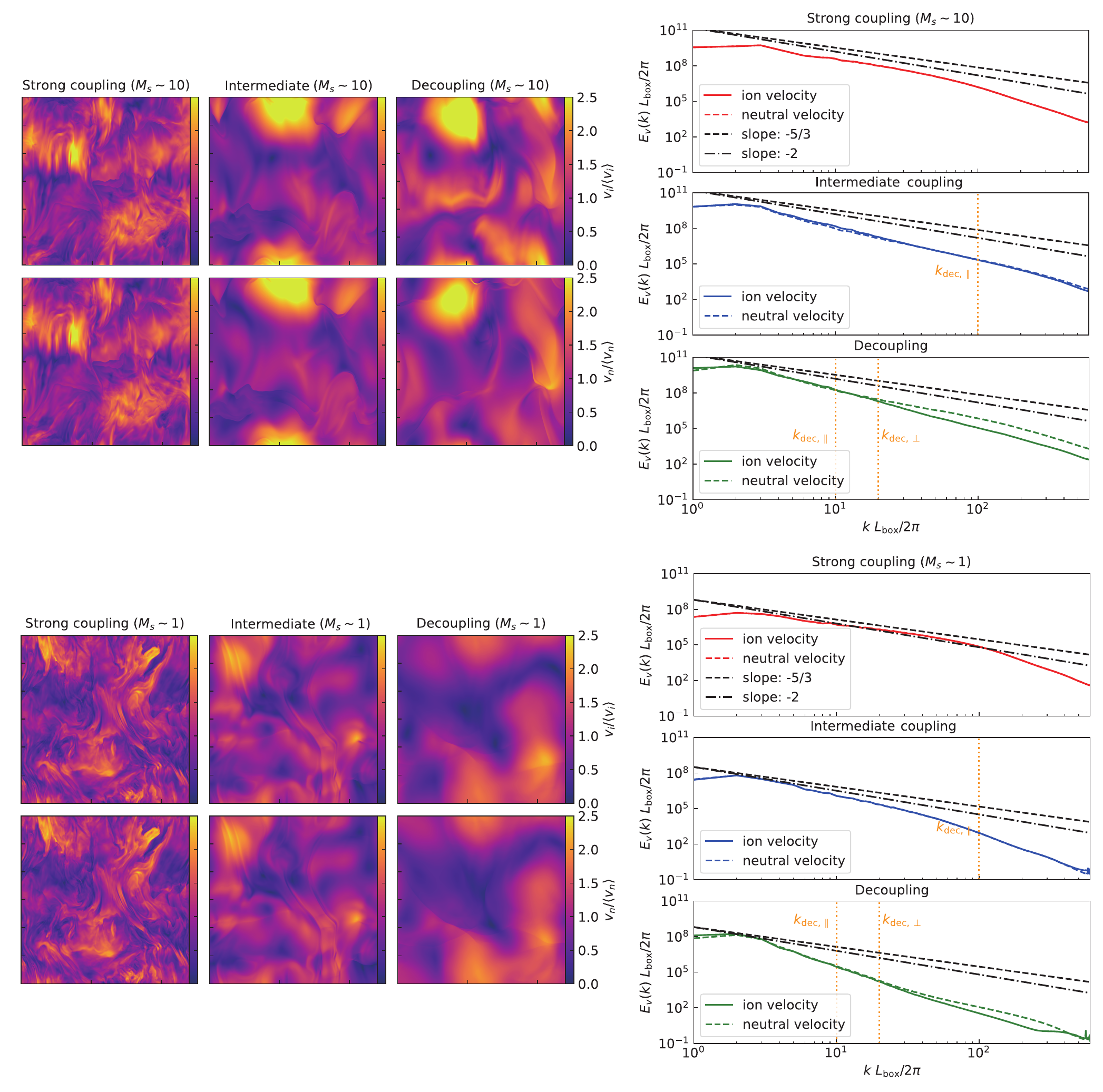}
        \caption{Left three panels: $x-z$ slices of ion velocity ($v_i$; top) and neutral velocity ($v_n$; bottom) at three different neutral-ion coupling statuses: strong coupling, intermediate coupling, and decoupling. Right panel: the kinetic energy spectrum. The theoretically expected parallel and perpendicular components of the neutral-ion decoupling wavenumber are denoted as $k_{\rm dec, \parallel}$ and $k_{\rm dec, \bot}$, respectively. To guide the eye, the dashed and dash-dotted black lines represent power-law slopes of -5/3 and -2, for comparison with Kolmogorov and Burgers turbulence scaling, respectively.}
    \label{fig:v_map}
\end{figure*}

\begin{figure*}[p]
\centering
\includegraphics[width=0.99\linewidth]{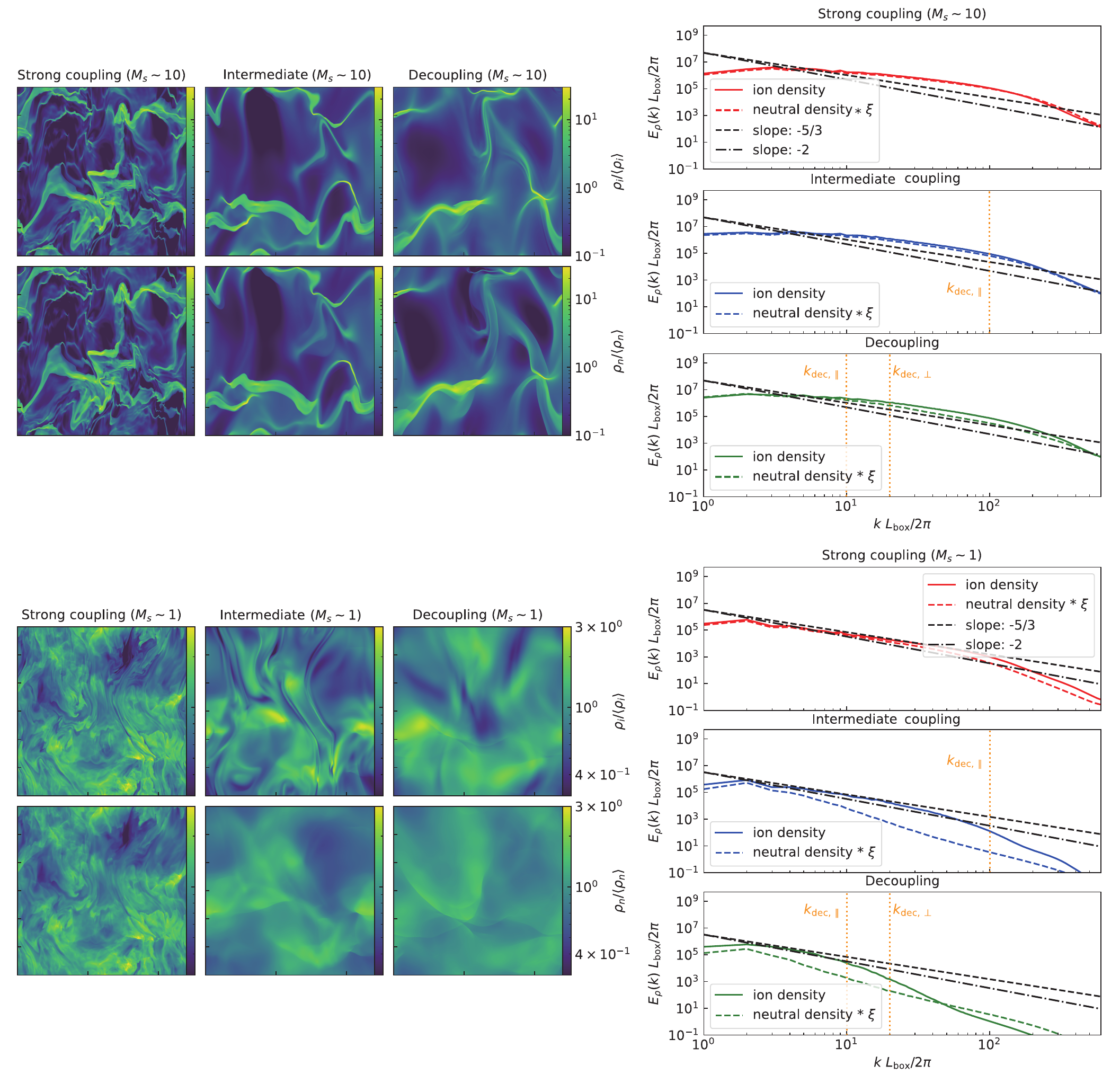}
        \caption{Left three panels:  $x-z$ slices of ion density ($\rho_i$; top) and neutral density ($\rho_n$; bottom) at three different neutral-ion coupling statuses: strong coupling, intermediate coupling, and decoupling. Right panel: the density spectrum. The neutral-ion decoupling wavenumber's theoretically expected parallel and perpendicular components are denoted as $k_{\rm dec, \parallel}$ and $k_{\rm dec, \bot}$, respectively. To guide the eye, the dashed and dash-dotted black lines represent power-law slopes of -5/3 and -2, for comparison with Kolmogorov and Burgers turbulence scaling, respectively.}
    \label{fig:rho_map}
\end{figure*}

\section{Numerical methods} 
\label{sec:data}
\subsection{Two-fluid simulations}
The scale-free 3D two-fluid simulations analyzed in this work are performed with the AthenaK code \citep{2024arXiv240916053S}. We model a two-fluid magneto-fluid system consisting of an ionized fluid (ions plus electrons) and a neutral fluid \citep{2024MNRAS.527.3945H}. The simulation domain is discretized on a 1200$^3$ uniformly staggered grid. Under the assumptions of an isothermal equation of state with constant sound speed $c_s$ and periodic boundary conditions, we solve the two-fluid MHD equations:
\begin{equation}
\label{eq.mhd}
\begin{aligned}
    &\partial\rho_i/\partial t +\nabla\cdot\left(\rho_i\mathbf{v}_i\right)=0,\\
    \\
    &\partial\rho_n/\partial t +\nabla\cdot(\rho_n\mathbf{v}_n)=0,\\ 
    \\
    &\partial(\rho_i\mathbf{v}_i)/\partial 
t+\nabla\cdot\left[\rho_i\mathbf{v}_i\mathbf{v}_i^T+\left(c_s^2\rho_i+\frac{B^2}{8\pi}\right)\mathbf{I}-\frac{\mathbf{B}\mathbf{B}^T}{4\pi}\right ]\\
    &~~~~~~~~~~~~~~~~~~~~~~~~~~~~~~~~~~~~~=\gamma_{\rm d}\rho_n\rho_i(\mathbf{v}_n-\mathbf{v}_i)+\mathbf{f}_i,\\     
    \\
    &\partial(\rho_n\mathbf{v}_n)/\partial t+\nabla\cdot\left[\rho_n\mathbf{v}_n\mathbf{v}_n^T+c_s^2\rho_n\mathbf{I}\right]
    =\gamma_{\rm d}\rho_n\rho_i(\mathbf{v}_i-\mathbf{v}_n)+\mathbf{f}_n,\\
    \\
&\partial\mathbf{B}/\partial t-\nabla\times(\mathbf{v}_i\times\mathbf{B})=0,\\
    \\
    &\nabla \cdot\mathbf{B}=0,
    \end{aligned}
\end{equation}
where $\rho$ and $\mathbf{v}$ denote the mass density and velocity of the ionized (subscript 
"$i$") and neutral (subscript "$n$") fluids, respectively, and $\mathbf{B}$ is the magnetic field. $\gamma_{\rm d}$ is the drag coefficient, while the stochastic forcing terms $\mathbf{f}_i$ and $\mathbf{f}_n$  drive turbulence in both fluids with weights proportional to their densities.
  
Initially, the ion and neutral densities are uniform, and the magnetic field is uniform along the $z$-axis. The ionization fraction is set to $\xi_i=\rho_i/(\rho_i+\rho_n)=0.1$. We generate two sets of simulations: (1) sonic Mach number $M_s \approx 1$ and Alfv\'en Mach number $M_A \approx 1$, and (2) $M_s \approx 10$ and $M_A \approx 1$, fully entering the strong turbulence regime. These are associated with physical conditions representing a 10~pc molecular cloud and a 0.01~pc dense clump, respectively. Specifically, we assume a velocity dispersion $v_{\rm inj}\approx 5$ km~s$^{-1}$ at 100~pc, Kolomogrov-type turbulence, and a temperature 10~K (giving $c_s\approx 0.2$~km~s$^{-1}$), which implies $v_{\rm inj}\approx 2$ and 0.2 km~s$^{-1}$ in the two simulations. These conditions correspond to (1) number density $n \approx 100$ cm$^{-3}$ and $B\approx \SI{15}{\micro G}$  for the 10~pc cloud and (2) $n \approx 10^5$ cm$^{-3}$ and $B\approx \SI{150}{\micro G}$ for the 0.01~pc clump, in agreement with existing Zeeman measurements \citep{2012ARA&A..50...29C}. To calculate the theoretically expected parallel and perpendicular wavenumber $k_{\rm dec, \parallel}$ and $k_{\rm dec, \bot}$, we adopt the mean values of density, magnetic field, and $v_{\rm inj}$.
Details of the simulations are listed in Tab.~\ref{tab:sim}.

\begin{figure*}
\centering
\includegraphics[width=0.99\linewidth]{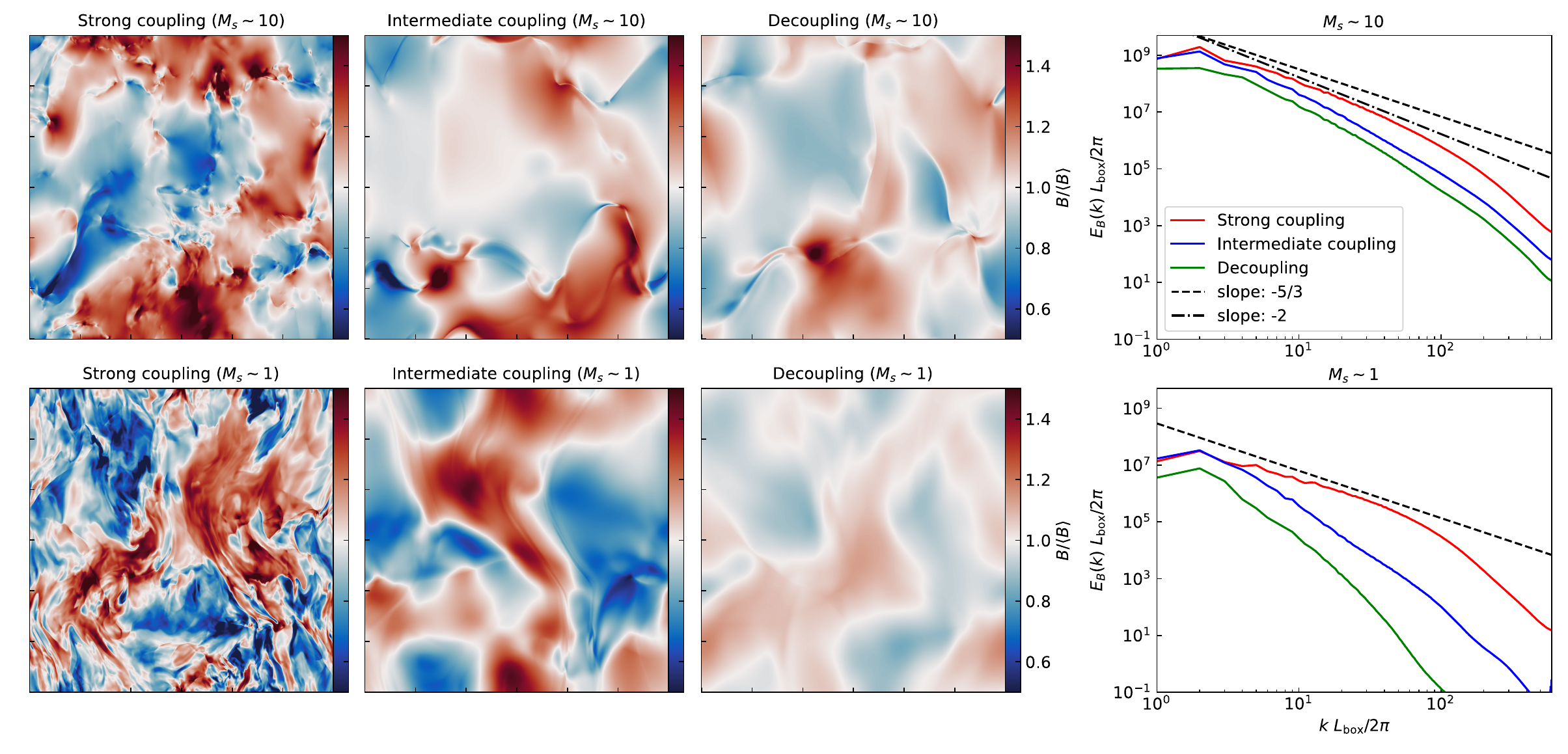}
        \caption{Left three panels:  $x-z$ slices of magnetic field strength at three different neutral-ion coupling statuses: strong coupling, intermediate coupling, and decoupling. Right panel: the magnetic field spectrum. To guide the eye, the dashed and dash-dotted black lines represent power-law slopes of -5/3 and -2, respectively.}
    \label{fig:B_map}
\end{figure*}


\begin{figure}
\centering
\includegraphics[width=0.99\linewidth]{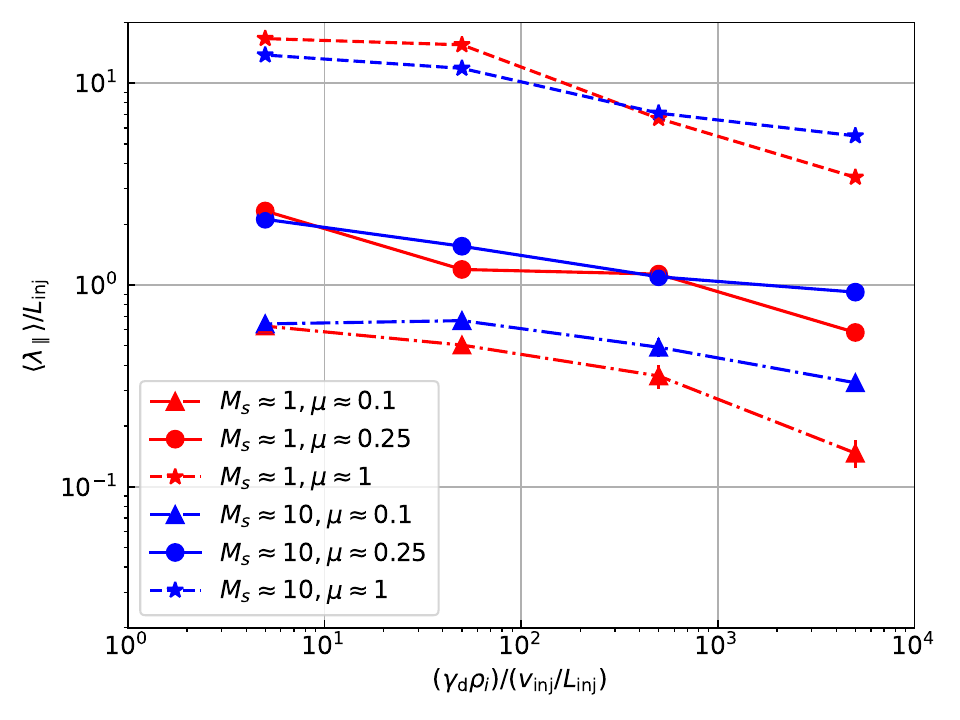}
        \caption{The parallel mean free path $\langle \lambda_\parallel\rangle$ in the unit of turbulence injection scale $L_{\rm inj}$ as a function of coupling strength $(\gamma_{\rm d} \rho_i) / (v_{\rm inj} / L_{\rm inj})$, which is the ratio of the neutral-ion collisional frequency to the eddy turnover rate at injection scale.}
    \label{fig:lambda}
\end{figure}
\section{Results} 
\label{sec:results}
\subsection{Damping of velocity, density, and magnetic field fluctuations}
In this section, we focus on studying the spatial distributions of velocity, density, and magnetic field, as well as their spectra. Their probability distribution functions are given in the Appendix.~\ref{appendix}.

\subsubsection{Velocity distribution and kinetic energy spectrum}
\textbf{Strongly coupled regime:} Fig.~\ref{fig:v_map}, corresponding to the simulations A1, A3, A4, B1, B3, and B4, presents 2D slices of the velocity fields and turbulent kinetic energy spectra for ions and neutrals. The slices are taken perpendicular to the mean magnetic field at the center of the simulation box. In both transonic and supersonic conditions, the velocity structures of ions (upper panel) and neutrals (lower panel) are highly similar in the strongly coupled regime, closely resembling the single-fluid approximation. For the transsonic case, the ion and neutral velocity spectra approximately follow the Kolmogorov scaling, exhibiting a spectral slope of -5/3. In contrast, the supersonic case yields steeper spectra with slopes close to -2, indicative of Burgers scaling. These steeper spectra result from energy dissipation associated with shocks.

\textbf{Intermediate coupling:} In the intermediate coupling regime, the 2D velocity slices and spectra of ions and neutrals remain similar to each other, but both spectra become steeper than the Kolmogorov and Burgers scalings. This steepening indicates significant damping of the turbulent fluctuations. In this regime, neutrals begin to decouple from ions in some local regions, although the decoupling is incomplete due to the fluctuations in ionization fractions. Consequently, neutrals cannot fully establish an independent hydrodynamic cascade, resulting in persistently steep spectra. The reduced value of $\gamma_{\rm d}$ and a high ionization fraction of 0.1 may cause enhanced ion inertial and frictional damping. However, a low ionization fraction 0.01 test is given in the Appendix in \cite{2024MNRAS.527.3945H}, showing that the neutral kinetic energy spectrum remains steep in the intermediate coupling regime. This behavior suggests that neutral-ion decoupling occurs gradually over a range of scales rather than sharply at a specific scale due to the spatial variation of ionization fraction. Compared to the strongly coupled regime, the velocity distributions of both neutrals and ions exhibit a clear deficiency of small-scale structures.

In addition to the reason that fluctuations in ionization fraction result in different local decoupling scales, this regime is also likely to have a contribution from the anisotropy of MHD turbulence, which means Alfv\'en waves propagate predominantly along the magnetic field, while the turbulent cascade develops preferentially in the perpendicular direction \citep{GS95,LV99}. As a result, neutrals may decouple from ions along the parallel direction yet remain coupled in the perpendicular direction due to turbulent motion. This behavior implies that neutral-ion coupling has both parallel and perpendicular components. Consequently, the decoupling scale is anisotropic, with its parallel component potentially much larger than the perpendicular component \citep{2015ApJ...810...44X,2024MNRAS.527.3945H}. In contrast, the classical model of neutral-ion coupling (see Eq.~\ref{eq.cp}) considers only Alfv\'en waves and thereby omits the perpendicular coupling.

\textbf{Decoupled regime:} In the decoupled regime, the ion and neutral spectra remain steep up to a wavenumber of approximately 30, which is close to the perpendicular decoupling scale. Beyond this scale, the neutral spectrum becomes shallower, indicating that neutrals have fully decoupled from ions and developed an independent hydrodynamic turbulent cascade characterized by a Kolmogorov slope. This trend was not observed in the earlier study by \cite{2024MNRAS.527.3945H} due to limited numerical resolution. Meanwhile, ions still experience frequent collisions with neutrals, effectively damping MHD turbulence in ions. We find that the damping is more severe in the transonic case with an even steeper spectra slope than that in the supersonic case. Likely, compressible slow and fast modes contribute differently in transonic and supersonic conditions, or shock locally alters ionization fraction, efficiently reducing the decoupling. Although the velocity slices of ions and neutrals exhibit noticeable differences, both species predominantly retain only large-scale fluctuations.

\subsubsection{Density distribution and density spectrum}
Fig.~\ref{fig:rho_map} presents 2D density slices and density spectra for ions and neutrals, for the simulations A1, A3, A4, B1, B3, and B4. Under supersonic conditions, the density structures of ions (upper panel) and neutrals (lower panel) remain broadly similar across all coupling regimes, although subtle differences emerge. As decoupling progresses, small-scale density structures become less pronounced. These density structures are more likely associated with turbulent cascade, which falls in different coupling regimes. In the strongly coupled regime, the neutral density spectra closely resemble those of ions, with both spectra exhibiting slopes shallower than the Kolmogorov scaling due to shock-enhanced small-scale density fluctuations. As the coupling weakens, the spectral slopes gradually steepen, but only slightly, as the shock compression does not suffer from damping. This steepening is more apparent for neutrals, reflecting increased damping of small-scale density fluctuations.

In contrast, under transonic conditions, where density fluctuations are primarily generated by the turbulent cascade, the density spectra exhibit distinct characteristics. In the strongly coupled regime, both ion and neutral density spectra generally follow a Kolmogorov-like scaling, consistent with their velocity spectra. However, as decoupling sets in, the ion density structures diverge markedly from those of the neutrals, despite the continued similarity of their velocity fields. The ion density distribution develops pronounced small-scale striations that are absent in the neutral counterpart. Correspondingly, the ion density spectrum remains close to the Kolmogorov scaling, whereas the neutral density spectrum steepens substantially.

This difference likely arises because the velocity field is dominated by incompressible Alfv\'enic turbulence, while density fluctuations are primarily driven by compressible turbulent motions. Although neutrals remain coupled to Alfv\'enic modes in the intermediate regime, their coupling to compressible MHD turbulence weakens, suppressing the formation of small-scale density structures associated with compressive motions. It is also possible that, when neutrals decouple from ions, the plasma $\beta$ of the ion fluid decreases. In this case, the sudden excess of magnetic fluctuation energy in the ion component can be converted into turbulent kinetic energy, further enhancing ion density fluctuations. While, over the relevant length scales, both magnetic and turbulent energy spectra steepen due to damping, the density fluctuations generated by the change in plasma $\beta$ can persist, allowing the ion density spectrum to retain a Kolmogorov-like form even in the intermediate and weakly coupled regimes. In the fully decoupled regime, filamentary structures in the ion density vanish, and the ion density spectrum steepens sharply beyond a wavenumber of approximately 10. The differences between the supersonic and transonic cases are most likely attributable to shocks: in supersonic turbulence, shock compression enhances density contrasts and thereby strengthens ion–neutral coupling.

\begin{figure*}
  \centering
  \gridline{
    \fig{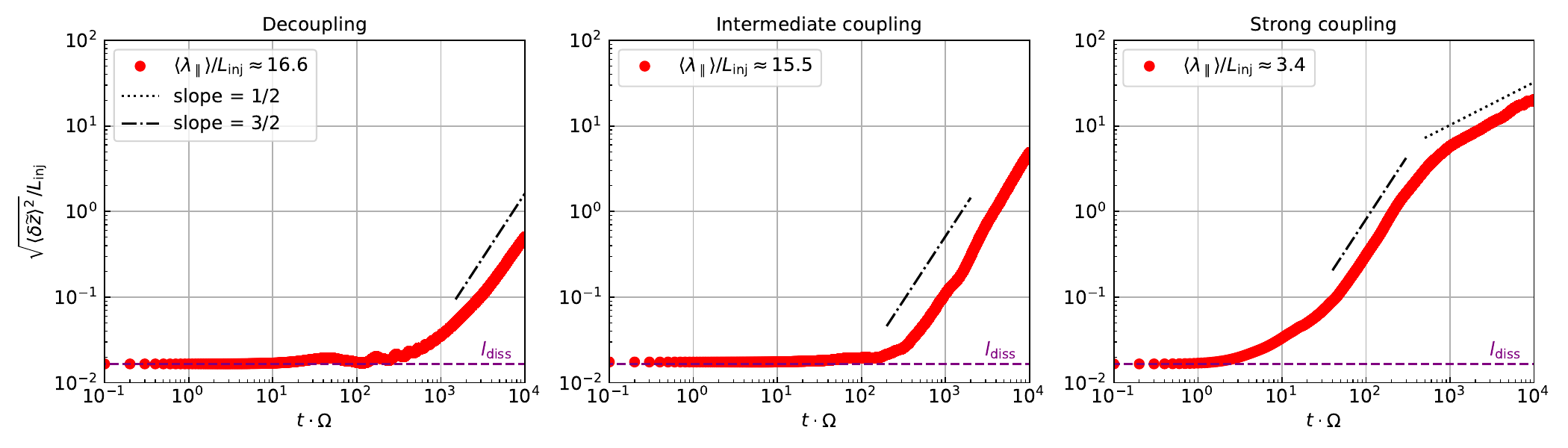}{0.99\textwidth}{(a) Initial $\mu\approx1$}
  }
  \vspace{-1.5em}
  \gridline{
    \fig{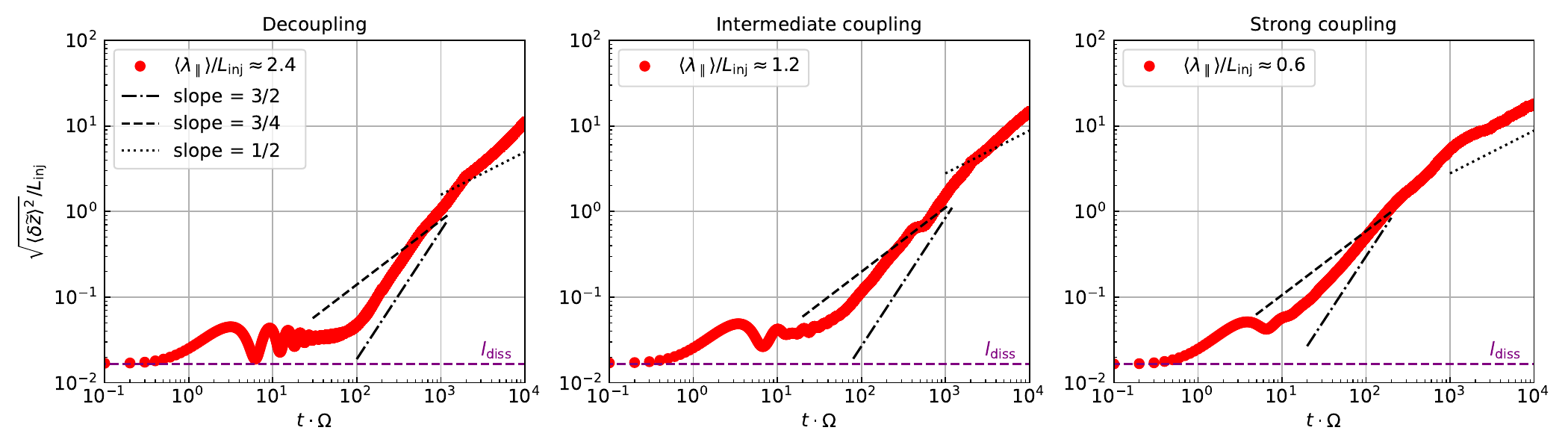}{0.99\textwidth}{(b) Initial $\mu\approx0.25$}
  }
  \vspace{-1.5em}
  \gridline{
    \fig{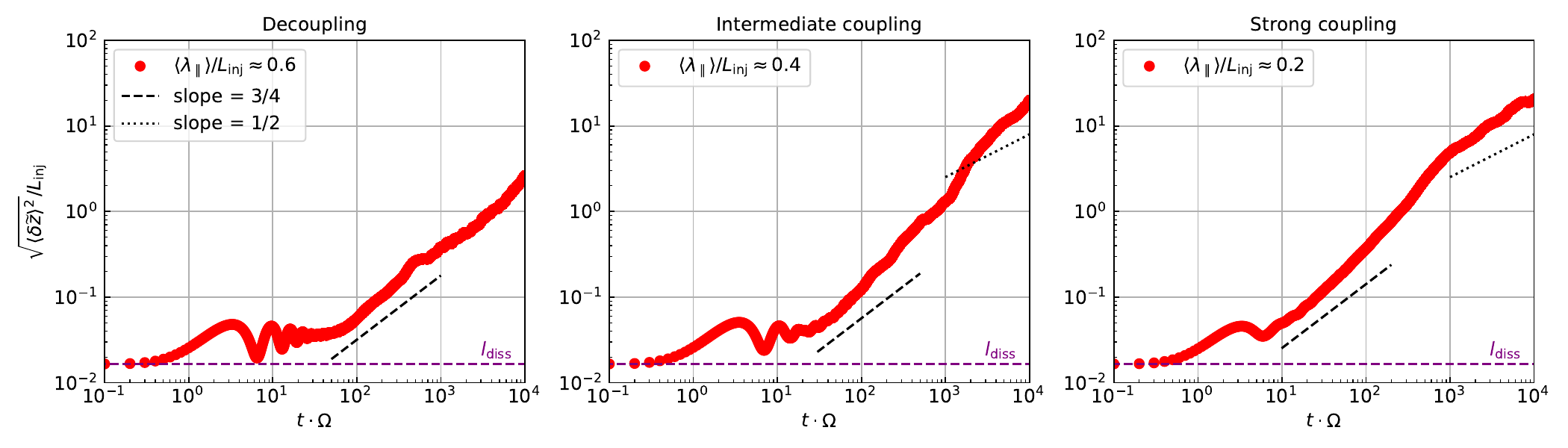}{0.99\textwidth}{(c) Initial $\mu\approx0.1$}
  }
  \caption{Plots of perpendicular displacement of the particles $\sqrt{\langle \delta \widetilde{z}\rangle^2}/L_{\rm inj}$ versus the CRs' gyro periods $t\cdot\Omega$ in transonic $M_s\approx1$ conditions, corresponding to the simulations A1, A3, and A4. $l_{\rm diss}$ is the numerical dissipation scale and $L_{\rm inj}$ is the injection scale. The panels (a), (b), and (c) correspond to initial cosine of pitch angle $\mu\approx1$, $\mu\approx0.25$, and $\mu\approx0.1$, respectively.}
  \label{fig:diff}
\end{figure*}

\subsubsection{Magnetic field distribution and magnetic field spectrum}
Fig.~\ref{fig:B_map} presents 2D slices of the total magnetic field strength and corresponding magnetic energy spectra. In the strongly coupled regime, magnetic field fluctuations exhibit filamentary structures under transonic conditions, closely resembling the velocity structures shown in Fig.~\ref{fig:v_map}. The magnetic energy spectrum in this case generally follows a Kolmogorov scaling with a slope of approximately -5/3. Under supersonic conditions, numerous discontinuities associated with shocks are evident, resulting in a steeper magnetic energy spectrum characterized by a slope close to -2. Small-scale magnetic field structures become significantly less prominent in the intermediate coupled and decoupled regimes. Correspondingly, the magnetic energy spectra steepen beyond a slope of -2 for both supersonic and transonic conditions, indicating that magnetic energy becomes increasingly concentrated at larger scales. This behavior naturally arises due to substantial damping caused by neutral-ion collisions, effectively suppressing magnetic fluctuations associated with the turbulent cascade.

\subsection{Examine the damping effect on CR's parallel mean free path}
Damping caused by neutral-ion collisions effectively suppresses magnetic fluctuations at smaller scales, implying reduced pitch-angle scattering of CRs and consequently a larger mean free path. To quantify the mean free path of CR particles, we adopt the methodology described in \cite{2013ApJ...779..140X}. Specifically, we inject 1000 test particles at random initial positions, assigning initial pitch angle cosine ($\mu$) around 1, 0.25, and 0.1, respectively. The particle Larmor radius $r_L = v/\Omega$ is set to 10 cells,  where $\Omega$ is the particle's gyrofrequency. We trace the change of pitch angle until it has changed by 90 degrees. The corresponding averaged distance traveled along the local magnetic field is taken as the parallel mean free path.


In Fig.~\ref{fig:lambda}, we present the mean free path $\langle\lambda_\parallel\rangle$ average over all particles as functions of the dimensionless coupling strength $(\gamma_{\rm d} \rho_i) / (v_{\rm inj} / L_{\rm inj})$, defined as the ratio of the neutral-ion collisional frequency to the eddy turnover rate at injection scale. 

In the strong-coupling regime, for CRs with initially large $\mu$, i.e., small pitch angles, the parallel motion of CRs along magnetic field lines is initially governed by the resonant pitch-angle scattering and undergoes scattering diffusion until the pitch angle becomes sufficiently large, and they enter the mirror diffusion regime. By contrast, for CRs with initially small $\mu$, i.e., large pitch angles, mirror diffusion is dominant, and $\langle \lambda_\parallel\rangle$ is determined by the characteristic size of magnetic mirrors with which mirroring particles predominantly interact \citep{2021ApJ...923...53L}. As mirroring of CRs effectively reverses their direction of motion along magnetic field lines, mirror diffusion strongly enhances CR confinement, resulting in a reduced mean free path at initially small $\mu$ \citep{2023ApJ...959L...8Z}. 

Nevertheless, under both supersonic and transonic conditions, $\langle\lambda_\parallel\rangle$ increases with decreasing coupling strength, consistent with stronger damping of MHD turbulence. Damping reduces pitch-angle scattering efficiency by suppressing magnetic fluctuations at gyroradius scales. However, the resulting increase in $\langle\lambda_\parallel\rangle$ is not significant—less than an order of magnitude—indicating that even with substantial damping and suppression of small-scale fluctuations, CRs at large pitch angles remain well confined by non-resonant mirror diffusion. In this regime, $\langle \lambda_\parallel\rangle$ can remain smaller than $L_{\rm inj}$ and the cloud size, ensuring that CRs are dynamically coupled to the cold ISM.

\subsection{CRs' perpendicular superdiffusion}
To investigate perpendicular superdiffusion, we follow the approach in \cite{2022MNRAS.512.2111H}, simultaneously injecting 50 beams of test particles randomly distributed within the simulation cube, with each beam containing 20 particles. The initial spatial separation between particles within each beam is set to 10 grid cells. Consistent with the previous setup, all particles have a Larmor radius of $r_L = 10$ cells and an initial $\mu\approx1$, $\mu\approx0.25$, or $\mu\approx0.1$, respectively. We trace particle trajectories along local magnetic field lines at each time step and interpolate these trajectories to measure the separation perpendicular to the mean magnetic field $\delta \widetilde{z}$ between particle trajectories at identical times. The root-mean-square (rms) value $\sqrt{\langle \delta \widetilde{z}\rangle^2}$ is taken as the perpendicular displacement of the particles.

Fig.~\ref{fig:diff} presents the time evolution of $\sqrt{\langle \delta \widetilde{z}\rangle^2}$ in transonic condition with initial $\mu\approx1$. In these conditions, the parallel mean free path $\langle\lambda_\parallel\rangle$ exceeds the turbulence injection scale $L_{\rm inj}$ across all three coupling regimes. After surpassing the numerical dissipation scale, $\sqrt{\langle \delta \widetilde{z}\rangle^2}$ exhibits a power-law growth with a slope of approximately 3/2. When
$\sqrt{\langle \delta \widetilde{z}\rangle^2}$ is larger than $L_{\rm inj}$, the growth slows down, showing a slope of approximately 1/2. The results in supersonic conditions are similar, as studied by \cite{2022MNRAS.512.2111H}. In the decoupled regime, however, $\sqrt{\langle \delta \widetilde{z}\rangle^2}$ grows more slowly before entering the perpendicular superdiffusion regime, requiring approximately 1000 gyroperiods compared to only 10 gyroperiods in the strongly coupled case. This delay arises because the initial perpendicular separation of CRs is smaller than the maximum parallel decoupling scale, and the damping of small-scale magnetic fluctuations modifies the superdiffusive behavior of magnetic field lines. The delay also makes it difficult to see the change of slope beyond $\sqrt{\langle \delta \widetilde{z}\rangle^2} > L_{\rm inj}$. As the required computational time increases exponentially by orders of magnitude, we stop the simulation at $10^4~~t\Omega$, while longer time may be investigated in future work.

We repeat this analysis for particles with initial $\mu\approx0.25$ and present the results in Fig.~\ref{fig:diff}.  In this scenario, mirror diffusion becomes more important, reducing $\langle\lambda_\parallel\rangle$. In the decoupled regime, where $\langle\lambda_\parallel\rangle>L_{\rm inj}$, the perpendicular separation $\sqrt{\langle \delta \widetilde{z}\rangle^2}$ again follows a power-law scaling with a slope of approximately 3/2. However, in the intermediate and strongly coupled regimes, where damping of MHD turbulence is less significant, the power-law slope gradually deviates from 3/2, approaching a slope closer to 3/4. The power-law slope does not perfectly align with 3/4 because trajectories are averaged over 1000 particles. Since $\langle\lambda_\parallel\rangle$ is not significantly smaller than $L_{\rm inj}$, some particles with larger mean free paths still contribute to a slope closer to 3/2. Although the perpendicular separation increases more slowly in these cases, the CR perpendicular motion remains superdiffusive. Additionally, $\sqrt{\langle \delta \widetilde{z}\rangle^2}$ exhibits oscillations before entering the superdiffusive regime. These oscillations, absent in the $\mu\approx1$ case, as shown in Fig.~\ref{fig:diff}, could result from the perpendicular displacement associated with the CRs' initial gyro-orbit. In the decoupled regime, small-scale magnetic fluctuations are damped, causing CRs to move along the magnetic field line ballistically, thus producing more pronounced oscillations in $\sqrt{\langle \delta \widetilde{z}\rangle^2}$. Nevertheless, when the initial $\mu\approx0.1$, the CRs' $\langle\lambda_\parallel\rangle$ is smaller than $L_{\rm inj}$ in the three coupling/decoupling regimes, and the oscillation possibly induced by mirrors is still apparent. The power-law slope of $\sqrt{\langle \delta \widetilde{z}\rangle^2}$ gradually deviates from 3/2, but is closer to 3/4. This finding is similar to that in \cite{2023ApJ...959L...8Z}, who measured the perpendicular superdiffusion of CRs with initially large pitch angles in a single-fluid MHD turbulence simulation. It shows that large-pitch-angle CRs are well confined in turbulence, irrespective of the coupling state of neutrals to ions. 

\section{Conclusion} 
\label{sec:conclusion}
In this work, we have examined the effects of partially ionized, turbulent environments on CR transport using high-resolution, 3D two-fluid simulations. By explicitly accounting for ion-neutral interactions, we have shown that despite the ion-neutral collisional damping on small-scale turbulent magnetic fluctuations, the damping effect on CR transport is not significant. Our main findings can be summarized as follows:
\begin{enumerate}
    \item Neutral-ion decoupling and turbulence damping: As neutral-ion coupling weakens, we observe substantial damping of velocity, density, and magnetic field fluctuations at small scales, resulting in steeper turbulence kinetic spectra than classical Kolmogorov or Burgers scalings. Under supersonic conditions, large-scale shocks do not suffer damping, sustaining the small-scale density enhancements due to the shock compression, thereby influencing turbulence structure and coupling strength, which is defined as the ratio of neutral-ion collisional frequency and eddy turnover time.
    \item Impact on CR parallel propagation: In the strongly coupled regime, frequent scattering and mirror diffusion can limit the parallel mean free path. Mirror diffusion becomes increasingly significant at larger pitch angles, decreasing the parallel mean free path and improving CR confinement. In the intermediately coupled and fully decoupled regimes, although the damping of small-scale magnetic fluctuations reduces resonant pitch angle scattering, the change of parallel mean free path of CRs is less than an order of magnitude, due to their non-resonant interactions with turbulent mirrors. The results imply that large-pitch-angle CRs can be well confined in the cold ISM, such as molecular clouds.
    \item Perpendicular CR superdiffusion: We identify two distinct superdiffusive regimes of CR transport perpendicular to the mean magnetic field: a diffusive regime, where the CRs' parallel mean free path is smaller than the turbulence injection scale, and a ballistic-like regime, where the parallel mean free path is larger than the injection scale.
    In the diffusive regime, CRs' perpendicular transport follows a superdiffusive scaling of $\propto t^{3/4}$. Conversely, in the ballistic-like regime, CRs' perpendicular diffusion is characterized by a scaling of $\propto t^{3/2}$. 
\end{enumerate}

\begin{acknowledgments}
Y.H. thanks Anatoly Spitkovsky for the helpful discussion. Y.H. acknowledges the support for this work provided by NASA through the NASA Hubble Fellowship grant No. HST-HF2-51557.001 awarded by the Space Telescope Science Institute, which is operated by the Association of Universities for Research in Astronomy, Incorporated, under NASA contract NAS5-26555. S.X. acknowledges the support from the NASA ATP award 80NSSC24K0896. A.L. acknowledges the support of NSF grants AST 2307840. This work used SDSC Expanse CPU and NCSA Delta CPU through allocations PHY230032, PHY230033, PHY230091, PHY230105,  PHY230178, and PHY240183, from the Advanced Cyberinfrastructure Coordination Ecosystem: Services \& Support (ACCESS) program, which is supported by National Science Foundation grants \#2138259, \#2138286, \#2138307, \#2137603, and \#2138296. 
\end{acknowledgments}

%

\vspace{5mm}

\software{Python3 \citep{10.5555/1593511}
          }


\newpage
\appendix
\section{Histograms of velocity, density, and magnetic field}
\label{appendix}
\begin{figure*}
\centering
\includegraphics[width=0.99\linewidth]{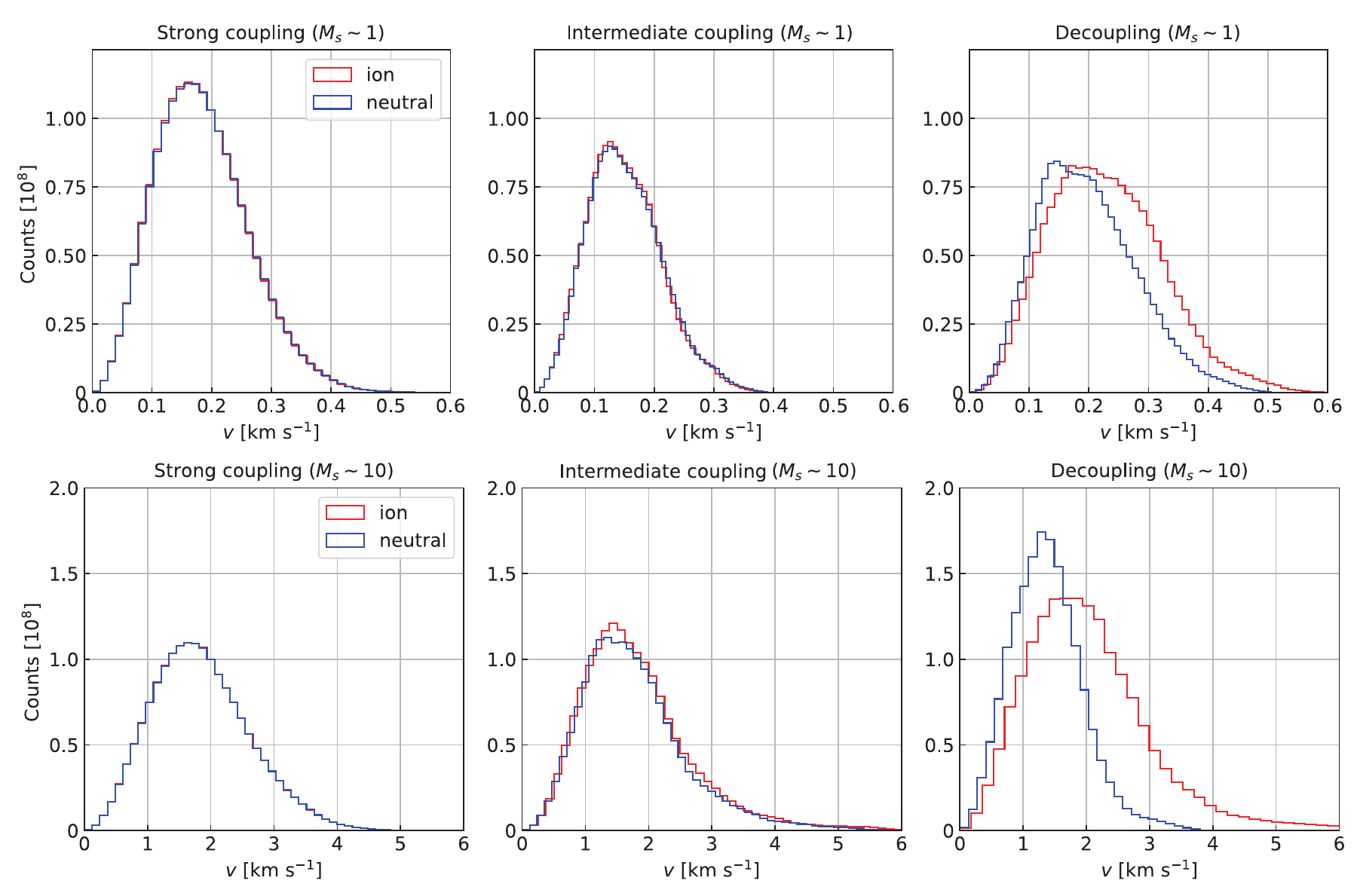}
        \caption{Histograms of ion velocity (red) and neutral velocity (blue) in transonic (top) and supersonic (bottom) conditions.}
    \label{fig:v_hist}
\end{figure*}

We present the ion and neutral velocity probability distribution functions (PDFs) in Fig.~\ref{fig:v_hist}. In the strong coupling regime, the PDFs for ions and neutrals are nearly identical. Differences become apparent in the intermediate coupling regime, with the largest divergence occurring in the decoupling region. Here, the ion velocity distribution shifts slightly toward higher values and broadens. This broadening is most likely caused by the unequal correlation time of the turbulence driving. Specifically, the initial correlation timescale of the driving force is set equal to the Alfv\'en speed crossing time in the strong coupling cases. When neutrals decouple from ions, they no longer support the development of Alfv\'en waves, and the effective Alfv\'en speed increases due to the reduced ion density. Consequently, the fixed correlation time in the simulation becomes too large for ions, resulting in a slower turbulence cascade and higher ion velocities \citep{2024MNRAS.527.3945H}. This effect is most pronounced in the supersonic decoupling case, where the sonic Mach numbers for neutrals and ions are approximately 7 and 11, respectively, compared to around 1.07 and 1.24 in the transonic decoupling case.

\begin{figure*}
\centering
\includegraphics[width=0.99\linewidth]{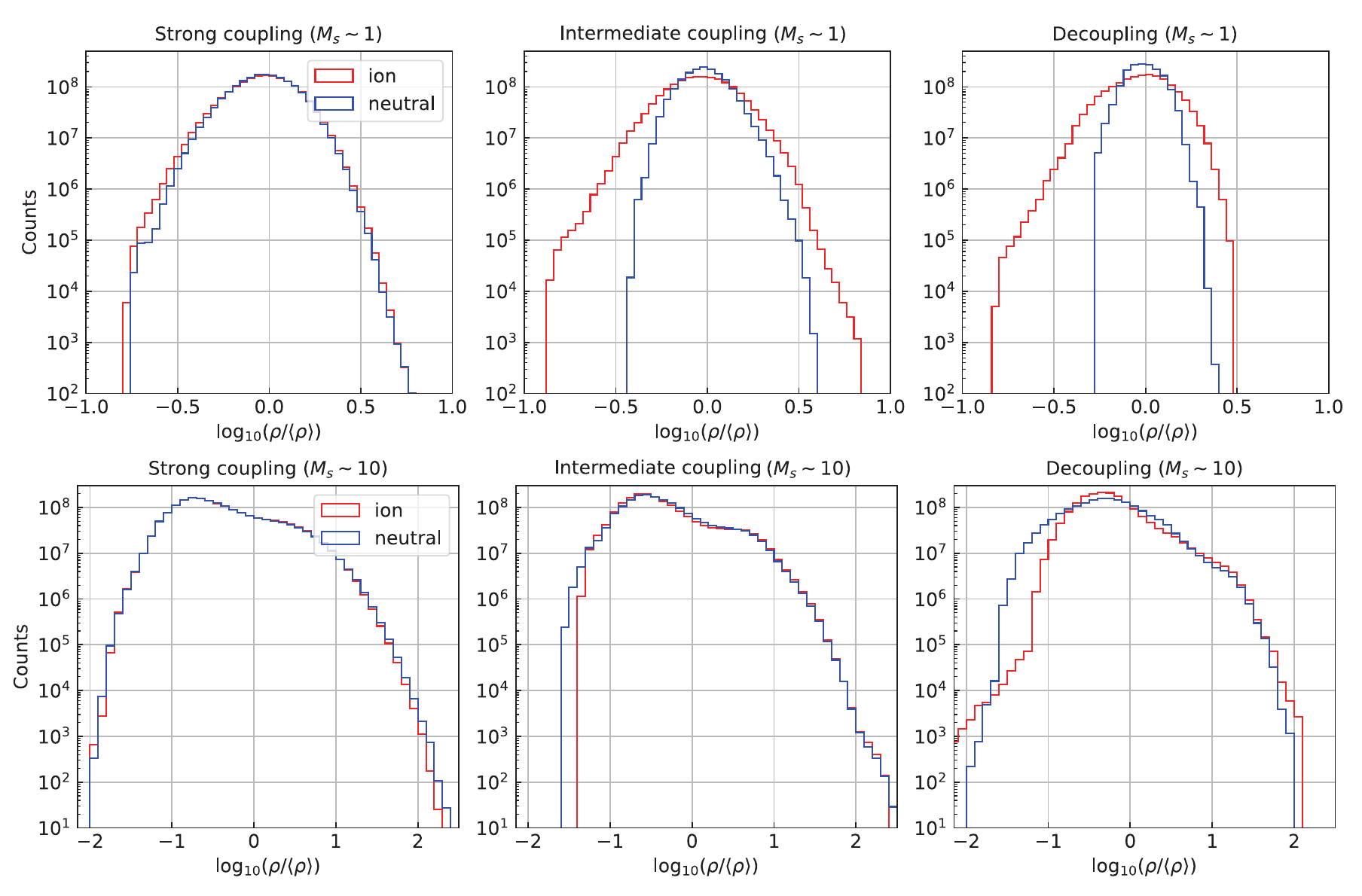}
        \caption{Same as Fig.~\ref{fig:v_hist}, but for the logarithmic mass density $\log_{10}(\rho/\langle \rho\rangle)$.}
    \label{fig:n_hist}
\end{figure*}

\begin{figure*}
\centering
\includegraphics[width=0.99\linewidth]{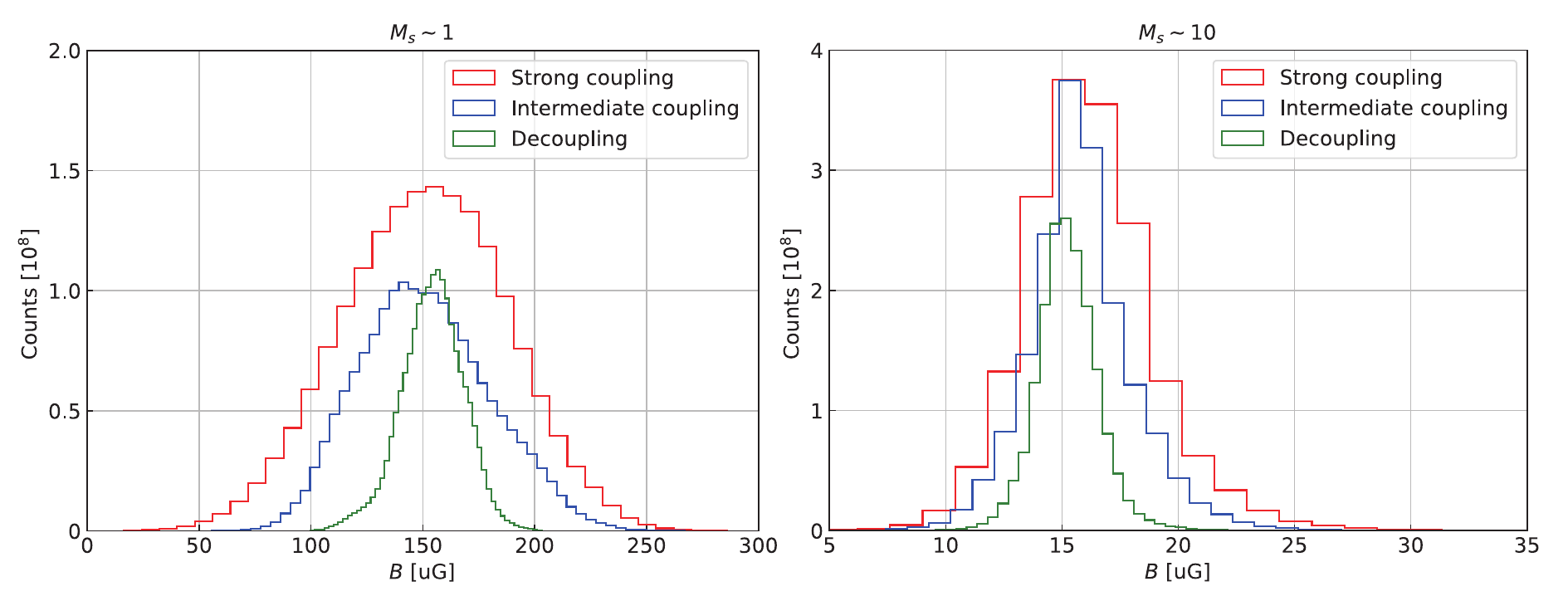}
        \caption{Histograms of magnetic field strength in transonic (left) and supersonic (right) conditions.}
    \label{fig:b_hist}
\end{figure*}

Fig.~\ref{fig:n_hist} shows the PDFs of the logarithmic mass densities of ions and neutrals, normalized by their respective mean values. Under supersonic conditions, the PDFs of ions and neutrals are similar, regardless of the coupling status, because shocks effectively enhance the coupling in the density field (see Fig.~\ref{fig:rho_map}). In transonic conditions with strong coupling, the ion PDFs closely resemble those of neutrals. However, in the intermediate coupling regime, the ion PDFs become more dispersed while the neutrals’ PDFs narrow, indicating more pronounced density fluctuations in ions. In the decoupling regimes, the PDFs of neutrals narrow further, signifying even less significant density fluctuations.

Fig.~\ref{fig:b_hist} presents the PDFs of the magnetic field strength. In both transonic and supersonic conditions, the PDFs in the strong coupling regime are broader than those in the decoupling regimes, indicating that magnetic field fluctuations are significantly damped when decoupling occurs.

\newpage
\bibliography{sample631}{}

\begin{thebibliography}{}
\expandafter\ifx\csname natexlab\endcsname\relax\def\natexlab#1{#1}\fi
\providecommand{\url}[1]{\href{#1}{#1}}
\providecommand{\dodoi}[1]{doi:~\href{http://doi.org/#1}{\nolinkurl{#1}}}
\providecommand{\doeprint}[1]{\href{http://ascl.net/#1}{\nolinkurl{http://ascl.net/#1}}}
\providecommand{\doarXiv}[1]{\href{https://arxiv.org/abs/#1}{\nolinkurl{https://arxiv.org/abs/#1}}}

\bibitem[{{Aharonian} {et~al.}(2008){Aharonian}, {Akhperjanian}, {Bazer-Bachi}, {Behera}, {Beilicke}, {Benbow}, {Berge}, {Bernl{\"o}hr}, {Boisson}, {Bolz}, {Borrel}, {Braun}, {Brion}, {Brown}, {B{\"u}hler}, {Bulik}, {B{\"u}sching}, {Boutelier}, {Carrigan}, {Chadwick}, {Chounet}, {Clapson}, {Coignet}, {Cornils}, {Costamante}, {Degrange}, {Dickinson}, {Djannati-Ata{\"\i}}, {Domainko}, {O'C. Drury}, {Dubus}, {Dyks}, {Egberts}, {Emmanoulopoulos}, {Espigat}, {Farnier}, {Feinstein}, {Fiasson}, {F{\"o}rster}, {Fontaine}, {Fukui}, {Funk}, {Funk}, {F{\"u}{\ss}ling}, {Gallant}, {Giebels}, {Glicenstein}, {Gl{\"u}ck}, {Goret}, {Hadjichristidis}, {Hauser}, {Hauser}, {Heinzelmann}, {Henri}, {Hermann}, {Hinton}, {Hoffmann}, {Hofmann}, {Holleran}, {Hoppe}, {Horns}, {Jacholkowska}, {de Jager}, {Kendziorra}, {Kerschhaggl}, {Kh{\'e}lifi}, {Komin}, {Kosack}, {Lamanna}, {Latham}, {Le Gallou}, {Lemi{\`e}re}, {Lemoine-Goumard}, {Lenain}, {Lohse}, {Martin}, {Martineau-Huynh}, {Marcowith}, {Masterson}, {Maurin}, {McComb}, {Moderski},
  {Moriguchi}, {Moulin}, {de Naurois}, {Nedbal}, {Nolan}, {Olive}, {Orford}, {Osborne}, {Ostrowski}, {Panter}, {Pedaletti}, {Pelletier}, {Petrucci}, {Pita}, {P{\"u}hlhofer}, {Punch}, {Ranchon}, {Raubenheimer}, {Raue}, {Rayner}, {Reimer}, {Renaud}, {Ripken}, {Rob}, {Rolland}, {Rosier-Lees}, {Rowell}, {Rudak}, {Ruppel}, {Sahakian}, {Santangelo}, {Saug{\'e}}, {Schlenker}, {Schlickeiser}, {Schr{\"o}der}, {Schwanke}, {Schwarzburg}, {Schwemmer}, {Shalchi}, {Sol}, {Spangler}, {Stawarz}, {Steenkamp}, {Stegmann}, {Superina}, {Takeuchi}, {Tam}, {Tavernet}, {Terrier}, {van Eldik}, {Vasileiadis}, {Venter}, {Vialle}, {Vincent}, {Vivier}, {V{\"o}lk}, {Volpe}, {Wagner}, \& {Ward}}]{2008A&A...481..401A}
{Aharonian}, F., {Akhperjanian}, A.~G., {Bazer-Bachi}, A.~R., {et~al.} 2008, \aap, 481, 401, \dodoi{10.1051/0004-6361:20077765}

\bibitem[{{Balsara}(1996)}]{1996ApJ...465..775B}
{Balsara}, D.~S. 1996, \apj, 465, 775, \dodoi{10.1086/177462}

\bibitem[{{Barreto-Mota} {et~al.}(2025){Barreto-Mota}, {de Gouveia Dal Pino}, {Xu}, \& {Lazarian}}]{2025ApJ...988..269B}
{Barreto-Mota}, L., {de Gouveia Dal Pino}, E.~M., {Xu}, S., \& {Lazarian}, A. 2025, \apj, 988, 269, \dodoi{10.3847/1538-4357/ade4c8}

\bibitem[{{Beattie} {et~al.}(2022){Beattie}, {Krumholz}, {Federrath}, {Sampson}, \& {Crocker}}]{2022FrASS...9.0900B}
{Beattie}, J.~R., {Krumholz}, M.~R., {Federrath}, C., {Sampson}, M.~L., \& {Crocker}, R.~M. 2022, Frontiers in Astronomy and Space Sciences, 9, 900900, \dodoi{10.3389/fspas.2022.900900}

\bibitem[{{Blasi}(2013)}]{2013A&ARv..21...70B}
{Blasi}, P. 2013, \aapr, 21, 70, \dodoi{10.1007/s00159-013-0070-7}

\bibitem[{{Burkhart} {et~al.}(2015){Burkhart}, {Lazarian}, {Balsara}, {Meyer}, \& {Cho}}]{2015ApJ...805..118B}
{Burkhart}, B., {Lazarian}, A., {Balsara}, D., {Meyer}, C., \& {Cho}, J. 2015, \apj, 805, 118, \dodoi{10.1088/0004-637X/805/2/118}

\bibitem[{{Butsky} {et~al.}(2020){Butsky}, {Fielding}, {Hayward}, {Hummels}, {Quinn}, \& {Werk}}]{2020ApJ...903...77B}
{Butsky}, I.~S., {Fielding}, D.~B., {Hayward}, C.~C., {et~al.} 2020, \apj, 903, 77, \dodoi{10.3847/1538-4357/abbad2}

\bibitem[{{Cao} {et~al.}(2025){Cao}, {Aharonian}, {Axikegu}, {Bai}, {Bao}, {Bastieri}, {Bi}, {Bi}, {Bian}, {Bukevich}, {Cao}, {Cao}, {Cao}, {Chang}, {Chang}, {Chen}, {Chen}, {Chen}, {Chen}, {Chen}, {Chen}, {Chen}, {Chen}, {Chen}, {Chen}, {Chen}, {Chen}, {Chen}, {Chen}, {Cheng}, {Cheng}, {Chu}, {Cui}, {Cui}, {Cui}, {Cui}, {Dai}, {Dai}, {Dai}, {Danzengluobu}, {Dong}, {Duan}, {Fan}, {Fan}, {Fang}, {Fang}, {Fang}, {Feng}, {Feng}, {Feng}, {Feng}, {Feng}, {Feng}, {Feng}, {Gabici}, {Gao}, {Gao}, {Gao}, {Gao}, {Gao}, {Ge}, {Ge}, {Geng}, {Giacinti}, {Gong}, {Gou}, {Gu}, {Guo}, {Guo}, {Guo}, {Guo}, {Guo}, {Han}, {Hannuksela}, {Hasan}, {He}, {He}, {He}, {He}, {Hor}, {Hou}, {Hou}, {Hou}, {Hu}, {Hu}, {Hu}, {Huang}, {Huang}, {Huang}, {Huang}, {Huang}, {Huang}, {Huang}, {Huang}, {Ji}, {Jia}, {Jia}, {Jiang}, {Jiang}, {Jiang}, {Jiang}, {Jin}, {Kang}, {Karpikov}, {Khangulyan}, {Kuleshov}, {Kurinov}, {Li}, {Li}, {Li}, {Li}, {Li}, {Li}, {Li}, {Li}, {Li}, {Li}, {Li}, {Li}, {Li}, {Li}, {Li}, {Li}, {Li}, {Li}, {Liang}, {Liang},
  {Lin}, {Liu}, {Liu}, {Liu}, {Liu}, {Liu}, {Liu}, {Liu}, {Liu}, {Liu}, {Liu}, {Liu}, {Liu}, {Liu}, {Liu}, {Luo}, {Luo}, {Lv}, {Ma}, {Ma}, {Ma}, {Mao}, {Min}, {Mitthumsiri}, {Mu}, {Nan}, {Neronov}, {Ng}, {Ou}, {Pattarakijwanich}, {Pei}, {Qi}, {Qi}, {Qiao}, {Qin}, {Raza}, {Ruffolo}, {S{\'a}iz}, {Saeed}, {Semikoz}, {Shao}, {Shchegolev}, {Sheng}, {Shu}, {Song}, {Stenkin}, {Stepanov}, {Su}, {Sun}, {Sun}, {Sun}, {Sun}, {Takata}, {Tam}, {Tang}, {Tang}, {Tang}, {Tian}, {Wan}, {Wang}, {Wang}, {Wang}, {Wang}, {Wang}, {Wang}, {Wang}, {Wang}, {Wang}, {Wang}, {Wang}, {Wang}, {Wang}, \& {Wang}}]{2025SCPMA..6879502C}
{Cao}, Z., {Aharonian}, F., {Axikegu}, {et~al.} 2025, Science China Physics, Mechanics, and Astronomy, 68, 279502, \dodoi{10.1007/s11433-024-2477-9}

\bibitem[{{Cesarsky} \& {Kulsrud}(1973)}]{1973ApJ...185..153C}
{Cesarsky}, C.~J., \& {Kulsrud}, R.~M. 1973, \apj, 185, 153, \dodoi{10.1086/152405}

\bibitem[{{Cho} {et~al.}(2002){Cho}, {Lazarian}, \& {Vishniac}}]{2002ApJ...564..291C}
{Cho}, J., {Lazarian}, A., \& {Vishniac}, E.~T. 2002, ApJ, 564, 291

\bibitem[{{Comisso} \& {Sironi}(2018)}]{2018PhRvL.121y5101C}
{Comisso}, L., \& {Sironi}, L. 2018, \prl, 121, 255101, \dodoi{10.1103/PhysRevLett.121.255101}

\bibitem[{{Comisso} \& {Sironi}(2019)}]{2019ApJ...886..122C}
---. 2019, \apj, 886, 122, \dodoi{10.3847/1538-4357/ab4c33}

\bibitem[{{Crutcher}(2012)}]{2012ARA&A..50...29C}
{Crutcher}, R.~M. 2012, \araa, 50, 29, \dodoi{10.1146/annurev-astro-081811-125514}

\bibitem[{{Draine}(1986)}]{1986MNRAS.220..133D}
{Draine}, B.~T. 1986, \mnras, 220, 133, \dodoi{10.1093/mnras/220.1.133}

\bibitem[{{Draine}(2011)}]{2011piim.book.....D}
---. 2011, {Physics of the Interstellar and Intergalactic Medium}

\bibitem[{{Everett} {et~al.}(2008){Everett}, {Zweibel}, {Benjamin}, {McCammon}, {Rocks}, \& {Gallagher}}]{2008ApJ...674..258E}
{Everett}, J.~E., {Zweibel}, E.~G., {Benjamin}, R.~A., {et~al.} 2008, \apj, 674, 258, \dodoi{10.1086/524766}

\bibitem[{{Eyink} {et~al.}(2013){Eyink}, {Vishniac}, {Lalescu}, {Aluie}, {Kanov}, {B{\"u}rger}, {Burns}, {Meneveau}, \& {Szalay}}]{2013Natur.497..466E}
{Eyink}, G., {Vishniac}, E., {Lalescu}, C., {et~al.} 2013, \nat, 497, 466, \dodoi{10.1038/nature12128}

\bibitem[{{Eyink} {et~al.}(2011){Eyink}, {Lazarian}, \& {Vishniac}}]{2011ApJ...743...51E}
{Eyink}, G.~L., {Lazarian}, A., \& {Vishniac}, E.~T. 2011, \apj, 743, 51, \dodoi{10.1088/0004-637X/743/1/51}

\bibitem[{{Giacalone} \& {Jokipii}(1999)}]{1999ApJ...520..204G}
{Giacalone}, J., \& {Jokipii}, J.~R. 1999, \apj, 520, 204, \dodoi{10.1086/307452}

\bibitem[{{Girichidis} {et~al.}(2016){Girichidis}, {Naab}, {Walch}, {Hanasz}, {Mac Low}, {Ostriker}, {Gatto}, {Peters}, {W{\"u}nsch}, {Glover}, {Klessen}, {Clark}, \& {Baczynski}}]{2016ApJ...816L..19G}
{Girichidis}, P., {Naab}, T., {Walch}, S., {et~al.} 2016, \apjl, 816, L19, \dodoi{10.3847/2041-8205/816/2/L19}

\bibitem[{{Goldreich} \& {Sridhar}(1995)}]{GS95}
{Goldreich}, P., \& {Sridhar}, S. 1995, \apj, 438, 763, \dodoi{10.1086/175121}

\bibitem[{{Hopkins} {et~al.}(2022{\natexlab{a}}){Hopkins}, {Butsky}, {Panopoulou}, {Ji}, {Quataert}, {Faucher-Gigu{\`e}re}, \& {Kere{\v{s}}}}]{2022MNRAS.516.3470H}
{Hopkins}, P.~F., {Butsky}, I.~S., {Panopoulou}, G.~V., {et~al.} 2022{\natexlab{a}}, \mnras, 516, 3470, \dodoi{10.1093/mnras/stac1791}

\bibitem[{{Hopkins} {et~al.}(2021){Hopkins}, {Chan}, {Ji}, {Hummels}, {Kere{\v{s}}}, {Quataert}, \& {Faucher-Gigu{\`e}re}}]{2021MNRAS.501.3640H}
{Hopkins}, P.~F., {Chan}, T.~K., {Ji}, S., {et~al.} 2021, \mnras, 501, 3640, \dodoi{10.1093/mnras/staa3690}

\bibitem[{{Hopkins} {et~al.}(2025){Hopkins}, {Quataert}, {Ponnada}, \& {Silich}}]{2025arXiv250118696H}
{Hopkins}, P.~F., {Quataert}, E., {Ponnada}, S.~B., \& {Silich}, E. 2025, arXiv e-prints, arXiv:2501.18696, \dodoi{10.48550/arXiv.2501.18696}

\bibitem[{{Hopkins} {et~al.}(2022{\natexlab{b}}){Hopkins}, {Squire}, {Butsky}, \& {Ji}}]{2022MNRAS.517.5413H}
{Hopkins}, P.~F., {Squire}, J., {Butsky}, I.~S., \& {Ji}, S. 2022{\natexlab{b}}, \mnras, 517, 5413, \dodoi{10.1093/mnras/stac2909}

\bibitem[{{Hu} {et~al.}(2021){Hu}, {Lazarian}, \& {Xu}}]{HLX21a}
{Hu}, Y., {Lazarian}, A., \& {Xu}, S. 2021, \apj, 915, 67, \dodoi{10.3847/1538-4357/ac00ab}

\bibitem[{{Hu} {et~al.}(2022){Hu}, {Lazarian}, \& {Xu}}]{2022MNRAS.512.2111H}
---. 2022, \mnras, 512, 2111, \dodoi{10.1093/mnras/stac319}

\bibitem[{{Hu} {et~al.}(2024){Hu}, {Xu}, {Arzamasskiy}, {Stone}, \& {Lazarian}}]{2024MNRAS.527.3945H}
{Hu}, Y., {Xu}, S., {Arzamasskiy}, L., {Stone}, J.~M., \& {Lazarian}, A. 2024, \mnras, 527, 3945, \dodoi{10.1093/mnras/stad3493}

\bibitem[{{IceCube Collaboration}(2013)}]{2013Sci...342E...1I}
{IceCube Collaboration}. 2013, Science, 342, 1242856, \dodoi{10.1126/science.1242856}

\bibitem[{{IceCube Collaboration} {et~al.}(2018){IceCube Collaboration}, {Aartsen}, {Ackermann}, {Adams}, {Aguilar}, {Ahlers}, {Ahrens}, {Samarai}, {Altmann}, {Andeen}, {Anderson}, {Ansseau}, {Anton}, {Arg{\"u}elles}, {Arsioli}, {Auffenberg}, {Axani}, {Bagherpour}, {Bai}, {Barron}, {Barwick}, {Baum}, {Bay}, {Beatty}, {Becker Tjus}, {Becker}, {BenZvi}, {Berley}, {Bernardini}, {Besson}, {Binder}, {Bindig}, {Blaufuss}, {Blot}, {Bohm}, {B{\"o}rner}, {Bos}, {B{\"o}ser}, {Botner}, {Bourbeau}, {Bourbeau}, {Bradascio}, {Braun}, {Brenzke}, {Bretz}, {Bron}, {Brostean-Kaiser}, {Burgman}, {Busse}, {Carver}, {Cheung}, {Chirkin}, {Christov}, {Clark}, {Classen}, {Coenders}, {Collin}, {Conrad}, {Coppin}, {Correa}, {Cowen}, {Cross}, {Dave}, {Day}, {de Andr{\'e}}, {De Clercq}, {DeLaunay}, {Dembinski}, {DeRidder}, {Desiati}, {de Vries}, {de Wasseige}, {de With}, {DeYoung}, {D{\'\i}az-V{\'e}lez}, {di Lorenzo}, {Dujmovic}, {Dumm}, {Dunkman}, {Dvorak}, {Eberhardt}, {Ehrhardt}, {Eichmann}, {Eller}, {Evenson}, {Fahey}, {Fazely},
  {Felde}, {Filimonov}, {Finley}, {Flis}, {Franckowiak}, {Friedman}, {Fritz}, {Gaisser}, {Gallagher}, {Gerhardt}, {Ghorbani}, {Giommi}, {Glauch}, {Gl{\"u}senkamp}, {Goldschmidt}, {Gonzalez}, {Grant}, {Griffith}, {Haack}, {Hallgren}, {Halzen}, {Hanson}, {Hebecker}, {Heereman}, {Helbing}, {Hellauer}, {Hickford}, {Hignight}, {Hill}, {Hoffman}, {Hoffmann}, {Hoinka}, {Hokanson-Fasig}, {Hoshina}, {Huang}, {Huber}, {Hultqvist}, {H{\"u}nnefeld}, {Hussain}, {In}, {Iovine}, {Ishihara}, {Jacobi}, {Japaridze}, {Jeong}, {Jero}, {Jones}, {Kalaczynski}, {Kang}, {Kappes}, {Kappesser}, {Karg}, {Karle}, {Katz}, {Kauer}, {Keivani}, {Kelley}, {Kheirandish}, {Kim}, {Kim}, {Kintscher}, {Kiryluk}, {Kittler}, {Klein}, {Koirala}, {Kolanoski}, {K{\"o}pke}, {Kopper}, {Kopper}, {Koschinsky}, {Koskinen}, {Kowalski}, {Krammer}, {Krings}, {Kroll}, {Kr{\"u}ckl}, {Kunwar}, {Kurahashi}, {Kuwabara}, {Kyriacou}, {Labare}, {Lanfranchi}, {Larson}, {Lauber}, {Leonard}, {Lesiak-Bzdak}, {Leuermann}, {Liu}, {Lozano Mariscal}, {Lu}, {L{\"u}nemann},
  {Luszczak}, {Madsen}, {Maggi}, {Mahn}, {Mancina}, {Maruyama}, {Mase}, {Maunu}, {Meagher}, {Medici}, {Meier}, {Menne}, {Merino}, {Meures}, {Miarecki}, {Micallef}, {Moment{\'e}}, {Montaruli}, {Moore}, {Morse}, {Moulai}, \& {Nahnhauer}}]{2018Sci...361..147I}
{IceCube Collaboration}, {Aartsen}, M.~G., {Ackermann}, M., {et~al.} 2018, Science, 361, 147, \dodoi{10.1126/science.aat2890}

\bibitem[{{Indriolo} {et~al.}(2007){Indriolo}, {Geballe}, {Oka}, \& {McCall}}]{2007ApJ...671.1736I}
{Indriolo}, N., {Geballe}, T.~R., {Oka}, T., \& {McCall}, B.~J. 2007, \apj, 671, 1736, \dodoi{10.1086/523036}

\bibitem[{{Jokipii}(1966)}]{1966ApJ...146..480J}
{Jokipii}, J.~R. 1966, \apj, 146, 480, \dodoi{10.1086/148912}

\bibitem[{{Jokipii} \& {Parker}(1969)}]{1969ApJ...155..777J}
{Jokipii}, J.~R., \& {Parker}, E.~N. 1969, \apj, 155, 777, \dodoi{10.1086/149909}

\bibitem[{{Jubelgas} {et~al.}(2008){Jubelgas}, {Springel}, {En{\ss}lin}, \& {Pfrommer}}]{2008A&A...481...33J}
{Jubelgas}, M., {Springel}, V., {En{\ss}lin}, T., \& {Pfrommer}, C. 2008, \aap, 481, 33, \dodoi{10.1051/0004-6361:20065295}

\bibitem[{{Kempski} {et~al.}(2023){Kempski}, {Fielding}, {Quataert}, {Galishnikova}, {Kunz}, {Philippov}, \& {Ripperda}}]{2023MNRAS.525.4985K}
{Kempski}, P., {Fielding}, D.~B., {Quataert}, E., {et~al.} 2023, \mnras, 525, 4985, \dodoi{10.1093/mnras/stad2609}

\bibitem[{{Kempski} \& {Quataert}(2022)}]{2022MNRAS.514..657K}
{Kempski}, P., \& {Quataert}, E. 2022, \mnras, 514, 657, \dodoi{10.1093/mnras/stac1240}

\bibitem[{{Krakau} \& {Schlickeiser}(2015)}]{2015ApJ...802..114K}
{Krakau}, S., \& {Schlickeiser}, R. 2015, \apj, 802, 114, \dodoi{10.1088/0004-637X/802/2/114}

\bibitem[{{Krumholz} {et~al.}(2023){Krumholz}, {Crocker}, \& {Offner}}]{2023MNRAS.520.5126K}
{Krumholz}, M.~R., {Crocker}, R.~M., \& {Offner}, S. S.~R. 2023, \mnras, 520, 5126, \dodoi{10.1093/mnras/stad459}

\bibitem[{{Kulsrud} \& {Pearce}(1969)}]{1969ApJ...156..445K}
{Kulsrud}, R., \& {Pearce}, W.~P. 1969, \apj, 156, 445, \dodoi{10.1086/149981}

\bibitem[{{Lazarian} \& {Vishniac}(1999)}]{LV99}
{Lazarian}, A., \& {Vishniac}, E.~T. 1999, \apj, 517, 700, \dodoi{10.1086/307233}

\bibitem[{{Lazarian} \& {Xu}(2021)}]{2021ApJ...923...53L}
{Lazarian}, A., \& {Xu}, S. 2021, \apj, 923, 53, \dodoi{10.3847/1538-4357/ac2de9}

\bibitem[{{Lazarian} {et~al.}(2023){Lazarian}, {Xu}, \& {Hu}}]{2023FrASS..1054760L}
{Lazarian}, A., {Xu}, S., \& {Hu}, Y. 2023, Frontiers in Astronomy and Space Sciences, 10, 1154760, \dodoi{10.3389/fspas.2023.1154760}

\bibitem[{{Lazarian} \& {Yan}(2014)}]{2014ApJ...784...38L}
{Lazarian}, A., \& {Yan}, H. 2014, \apj, 784, 38, \dodoi{10.1088/0004-637X/784/1/38}

\bibitem[{{Lemoine}(2023)}]{2023JPlPh..89e1701L}
{Lemoine}, M. 2023, Journal of Plasma Physics, 89, 175890501, \dodoi{10.1017/S0022377823000946}

\bibitem[{{Marcowith}(2025)}]{2025FrASS..1111076M}
{Marcowith}, A. 2025, Frontiers in Astronomy and Space Sciences, 11, 1411076, \dodoi{10.3389/fspas.2024.1411076}

\bibitem[{{McKee} {et~al.}(2010){McKee}, {Li}, \& {Klein}}]{2010ApJ...720.1612M}
{McKee}, C.~F., {Li}, P.~S., \& {Klein}, R.~I. 2010, \apj, 720, 1612, \dodoi{10.1088/0004-637X/720/2/1612}

\bibitem[{{Qin} {et~al.}(2002){Qin}, {Matthaeus}, \& {Bieber}}]{2002ApJ...578L.117Q}
{Qin}, G., {Matthaeus}, W.~H., \& {Bieber}, J.~W. 2002, \apjl, 578, L117, \dodoi{10.1086/344687}

\bibitem[{{Ruszkowski} {et~al.}(2017){Ruszkowski}, {Yang}, \& {Zweibel}}]{2017ApJ...834..208R}
{Ruszkowski}, M., {Yang}, H. Y.~K., \& {Zweibel}, E. 2017, \apj, 834, 208, \dodoi{10.3847/1538-4357/834/2/208}

\bibitem[{{Shu}(1992)}]{1992pavi.book.....S}
{Shu}, F.~H. 1992, {The physics of astrophysics. Volume II: Gas dynamics.}

\bibitem[{{Stone} {et~al.}(2024){Stone}, {Mullen}, {Fielding}, {Grete}, {Guo}, {Kempski}, {Most}, {White}, \& {Wong}}]{2024arXiv240916053S}
{Stone}, J.~M., {Mullen}, P.~D., {Fielding}, D., {et~al.} 2024, arXiv e-prints, arXiv:2409.16053, \dodoi{10.48550/arXiv.2409.16053}

\bibitem[{{Tilley} \& {Balsara}(2010)}]{2010MNRAS.406.1201T}
{Tilley}, D.~A., \& {Balsara}, D.~S. 2010, \mnras, 406, 1201, \dodoi{10.1111/j.1365-2966.2010.16768.x}

\bibitem[{{Tilley} \& {Balsara}(2011)}]{2011MNRAS.415.3681T}
---. 2011, \mnras, 415, 3681, \dodoi{10.1111/j.1365-2966.2011.18982.x}

\bibitem[{Van~Rossum \& Drake(2009)}]{10.5555/1593511}
Van~Rossum, G., \& Drake, F.~L. 2009, Python 3 Reference Manual (Scotts Valley, CA: CreateSpace)

\bibitem[{{VERITAS Collaboration} {et~al.}(2009){VERITAS Collaboration}, {Acciari}, {Aliu}, {Arlen}, {Aune}, {Bautista}, {Beilicke}, {Benbow}, {Boltuch}, {Bradbury}, {Buckley}, {Bugaev}, {Byrum}, {Cannon}, {Celik}, {Cesarini}, {Chow}, {Ciupik}, {Cogan}, {Colin}, {Cui}, {Dickherber}, {Duke}, {Fegan}, {Finley}, {Finnegan}, {Fortin}, {Fortson}, {Furniss}, {Galante}, {Gall}, {Gibbs}, {Gillanders}, {Godambe}, {Grube}, {Guenette}, {Gyuk}, {Hanna}, {Holder}, {Horan}, {Hui}, {Humensky}, {Imran}, {Kaaret}, {Karlsson}, {Kertzman}, {Kieda}, {Kildea}, {Konopelko}, {Krawczynski}, {Krennrich}, {Lang}, {Lebohec}, {Maier}, {McArthur}, {McCann}, {McCutcheon}, {Millis}, {Moriarty}, {Mukherjee}, {Nagai}, {Ong}, {Otte}, {Pandel}, {Perkins}, {Pizlo}, {Pohl}, {Quinn}, {Ragan}, {Reyes}, {Reynolds}, {Roache}, {Rose}, {Schroedter}, {Sembroski}, {Smith}, {Steele}, {Swordy}, {Theiling}, {Thibadeau}, {Varlotta}, {Vassiliev}, {Vincent}, {Wagner}, {Wakely}, {Ward}, {Weekes}, {Weinstein}, {Weisgarber}, {Williams}, {Wissel}, {Wood}, \&
  {Zitzer}}]{2009Natur.462..770V}
{VERITAS Collaboration}, {Acciari}, V.~A., {Aliu}, E., {et~al.} 2009, \nat, 462, 770, \dodoi{10.1038/nature08557}

\bibitem[{{Xiao} {et~al.}(2025){Xiao}, {Zhang}, \& {Xu}}]{2025A&A...699A.317X}
{Xiao}, Y.-W., {Zhang}, J.-F., \& {Xu}, S. 2025, \aap, 699, A317, \dodoi{10.1051/0004-6361/202453340}

\bibitem[{Xu(2019)}]{xu2019study}
Xu, S. 2019, Study on Magnetohydrodynamic Turbulence and Its Astrophysical Applications, Springer Theses (Springer Nature Singapore).
\newblock \url{https://books.google.com/books?id=oXWUDwAAQBAJ}

\bibitem[{{Xu} {et~al.}(2015){Xu}, {Lazarian}, \& {Yan}}]{2015ApJ...810...44X}
{Xu}, S., {Lazarian}, A., \& {Yan}, H. 2015, \apj, 810, 44, \dodoi{10.1088/0004-637X/810/1/44}

\bibitem[{{Xu} \& {Li}(2023)}]{2023ApJ...957...97X}
{Xu}, S., \& {Li}, H. 2023, \apj, 957, 97, \dodoi{10.3847/1538-4357/acfca5}

\bibitem[{{Xu} \& {Yan}(2013)}]{2013ApJ...779..140X}
{Xu}, S., \& {Yan}, H. 2013, \apj, 779, 140, \dodoi{10.1088/0004-637X/779/2/140}

\bibitem[{{Xu} {et~al.}(2016){Xu}, {Yan}, \& {Lazarian}}]{2016ApJ...826..166X}
{Xu}, S., {Yan}, H., \& {Lazarian}, A. 2016, \apj, 826, 166, \dodoi{10.3847/0004-637X/826/2/166}

\bibitem[{{Yan} \& {Lazarian}(2008)}]{2008ApJ...673..942Y}
{Yan}, H., \& {Lazarian}, A. 2008, \apj, 673, 942, \dodoi{10.1086/524771}

\bibitem[{{Yusef-Zadeh} {et~al.}(2024){Yusef-Zadeh}, {Wardle}, {Arendt}, {Hewitt}, {Hu}, {Lazarian}, {Kassim}, {Hyman}, \& {Heywood}}]{2024MNRAS.527.1275Y}
{Yusef-Zadeh}, F., {Wardle}, M., {Arendt}, R., {et~al.} 2024, \mnras, 527, 1275, \dodoi{10.1093/mnras/stad3203}

\bibitem[{{Zhang} \& {Xu}(2023)}]{2023ApJ...959L...8Z}
{Zhang}, C., \& {Xu}, S. 2023, \apjl, 959, L8, \dodoi{10.3847/2041-8213/ad0fe5}

\bibitem[{{Zhang} \& {Xu}(2024)}]{2024ApJ...975...65Z}
---. 2024, \apj, 975, 65, \dodoi{10.3847/1538-4357/ad79fb}

\bibitem[{{Zweibel}(2013)}]{2013PhPl...20e5501Z}
{Zweibel}, E.~G. 2013, Physics of Plasmas, 20, 055501, \dodoi{10.1063/1.4807033}

\end{thebibliography}
\bibliographystyle{aasjournal}



\end{document}